\documentclass{emulateapj}

\newcommand{\chandra}{{\it Chandra}}
\newcommand{\swift}{{\it Swift}}
\newcommand{\xmm}{{\it XMM-Newton}}
\newcommand{\suzaku}{{\it Suzaku}}

\begin{document}

\markboth{J.-U. Ness et al.}{X-ray dips in U\,Sco}

\title{From X-ray dips to eclipse: Witnessing disk reformation in the recurrent nova U\,Sco}

\author{J.-U. Ness\altaffilmark{1},
 B.E. Schaefer\altaffilmark{2},
 A. Dobrotka\altaffilmark{3,4},
 A. Sadowski\altaffilmark{5},
 J.J. Drake\altaffilmark{6},
 R. Barnard\altaffilmark{6},
 A. Talavera\altaffilmark{1},
R. Gonzalez-Riestra\altaffilmark{1},
 K.L. Page\altaffilmark{7},
 M. Hernanz\altaffilmark{8},
 G. Sala\altaffilmark{9},
 S. Starrfield\altaffilmark{10}
}

\altaffiltext{1}{XMM-Newton Science Operations Centre, ESA, PO Box 78, 28691 Villanueva de la Ca\~nada, Madrid, Spain, juness@sciops.esa.int}
\altaffiltext{2}{Department of Physics and Astronomy, Louisiana State
University, Baton Rouge, Louisiana, 70803, USA}
\altaffiltext{3}{Department of Physics, Institute of Materials Science, Faculty of Materials Science and Technology, Slovak University of Technology, J\'ana Bottu 25, 91724 Trnava, The Slovak Republic}
\altaffiltext{4}{Department of Astronomy, Graduate School of Science, Kyoto University, Sakyo-ku, Kyoto 606-8502, Japan}
\altaffiltext{5}{N. Copernicus Astronomical Center, Polish Academy of Sciences, Bartycka 18, 00-716 Warszawa, Poland}
\altaffiltext{6}{Harvard-Smithsonian Center for Astrophysics, 60 Garden Street, Cambridge, MA 02138, USA}
\altaffiltext{7}{Department of Physics \& Astronomy, University of Leicester, Leicester, LE1 7RH, UK}
\altaffiltext{8}{Institut de Ci\`encies de l'Espai (CSIC-IEEC), Campus UAB, Facultat de Ci\`encies, C5 parell 2$^{on}$, 08193 Bellaterra (Barcelona), Spain}
\altaffiltext{9}{Departament F\'{\i}sica i Enginyeria Nuclear, EUETIB (UPC-IEEC), Comte d'Urgell 187, 08036 Barcelona, Spain}
\altaffiltext{10}{School of Earth and Space Exploration, Arizona State University, Tempe, AZ 85287-1404, USA}

\begin{abstract}
 The 10th recorded outburst of the recurrent eclipsing nova U\,Sco
was observed simultaneously in X-ray, UV, and optical by \xmm\ on
days 22.9 and 34.9 after outburst. Two full passages of the
companion in front of the nova ejecta were observed, witnessing the
reformation of the accretion disk.\\
On day 22.9, we observed smooth eclipses in UV and optical but
deep dips in the X-ray light curve which disappeared by day 34.9,
then yielding clean eclipses in all bands.
X-ray dips can be caused by clumpy absorbing material that
intersects the line of sight while moving along highly elliptical
trajectories. Cold material from the companion could explain
the absence of dips in UV and optical light.
The disappearance of X-ray dips before day 34.9 implies significant
progress in the formation of the disk.\\
The X-ray spectra contain photospheric continuum emission plus
strong emission lines, but no clear absorption lines.
Both continuum and emission lines in the X-ray spectra
indicate a temperature increase from day 22.9 to
day 34.9. We find clear evidence in the spectra and light curves for
Thompson scattering of the photospheric emission from the white
dwarf. Photospheric absorption lines can be smeared out
during scattering in a plasma of fast electrons. We also find
spectral signatures of resonant line scattering that lead to
the observation of the strong emission lines. Their dominance
could be a general phenomenon in high-inclination systems such
as Cal\,87.
\end{abstract}


\section{Introduction}

 A nova outburst is a thermonuclear explosion on the surface of
a white dwarf star. The energy is produced by fusion of
hydrogen to helium and is released via complex radiation transport
processes through the optically thick ejecta, spanning a broad
wavelength range. As the density of the ejecta decreases with time,
the peak of the radiation output continuously shifts from long
wavelength optical emission to shorter wavelengths. The decreasing
density allows us to see deeper into the outflow where the temperature
and thus the energy of the emitted radiation is higher.
 Nova explosions occur in binary systems because the required amount of
hydrogen-rich burning material can only be obtained by accretion
from a second star that orbits the white dwarf at a close enough
distance to allow mass transfer via an accretion disk, see e.g.,
\cite{robinson76}. Once all the previously accreted hydrogen is consumed,
the nova will turn off, gradually returning to the original
condition of the system. A nova outburst does not significantly
change the system parameters, and mass transfer can set in again,
leading to new accumulation of hydrogen-rich material. Typically,
the accumulation of sufficient hydrogen for another outburst is
expected to take $\sim 10^4-10^5$ years, except for
a handful of recurrent novae, in which this cycle occurs fast
enough that more than one outburst has been observed in
a human lifetime \citep{schaefer10}.

 During the initial blast, the accretion disk is destroyed, which was
recently shown for U\,Sco by \cite{drake10} utilizing hydrodynamic
models. One question of interest is how quickly it reforms. This is
not well known,
and opportunities of direct observations of the reformation process are rare.
 Significant theoretical efforts
have been undertaken to understand the formation of accretion
disks. An illustration of the expected formation process via Roche Lobe
overflow is shown in fig.~1 in a review paper by \cite{verbunt82}.
Initially, the gas streams with supersonic velocity along a highly
eccentric orbit. As the stream hits the incoming gases, the flow
downstream changes into a circular orbit, which is expected to occur
on time scales longer than radiative cooling and the dynamical or
orbital timescale. In most cases, viscosity
plays an important role for the details, but this fundamental picture
remains the same. More important should be the presence of intensive
high-energy radiation during the times of the \xmm\ observations of U\,Sco.
The recurrent outbursts in the
eclipsing system U\,Sco are a unique laboratory to observe the
earliest stages of the disk formation.

 U\,Sco is a recurrent nova with now 10 recorded outbursts since 1863.
The latest outburst was discovered by B. Harris on 2010 Jan 28.4385
($=$HJD\,2455224.94346) at $V=7.8$\,mag \citep{iauc9111}.
\cite{usco_discovery} extrapolated the light curve to earlier times,
indicating that the peak magnitude was $V=7.5$ on 2010 Jan 28.1
($=$JD\,2455224.605), and that the time of outburst was
\begin{equation}
 t_0={\rm 2010 Jan 27.8}\ (={\rm JD\,2455224.305}).
\end{equation}
 This time is used as the reference
time $t_0$ in this article. The underlying system was identified
as an eclipsing system by \cite{schaefer_eclipse} with an
inclination angle of $\stackrel{>}{_\sim}80^{\rm o}$
\citep{thoroughgood01}. Based on 29 accurate eclipse times from
2001-2009, we find a best fit linear ephemeris in units of days to be
\begin{equation}
\label{ephem}
 HJDmin = 2451234.5387 + E*1.23054695,
\end{equation}
 where $E$ is an
integer that counts the cycles.
The binary separation is $6.5\,\pm\,0.4$\,R$_\odot$ 
\citep{schaefer_eclipse}, and the component masses have been determined by 
\cite{thoroughgood01} to be $>1.31$\,M$_\odot$ for the white dwarf
and $0.88\,\pm\,0.17$\,M$_\odot$ for the secondary. The radius of
the companion is $2.1\,\pm\,0.2$\,R$_\odot$, indicating that it
is evolved \citep{schaefer_eclipse}. The high mass of the white dwarf,
close to the Chandrasekhar mass limit, is consistent with the short
recurrence time scale and led to speculations that U\,Sco may be a
supernova Ia progenitor \citep{starrf88}. However, recent abundance
measurements by \cite{mason2011} indicate that the underlying white
dwarf may of the ONeMg kind which does not contain enough nuclear
binding energy for a SN\,Ia exposion. Consequently, even if the
Chandrasekhar mass limit is reached, the white dwarf may turn
into a neutron star via core collapse.

\begin{figure}[!ht]
\resizebox{\hsize}{!}{\includegraphics{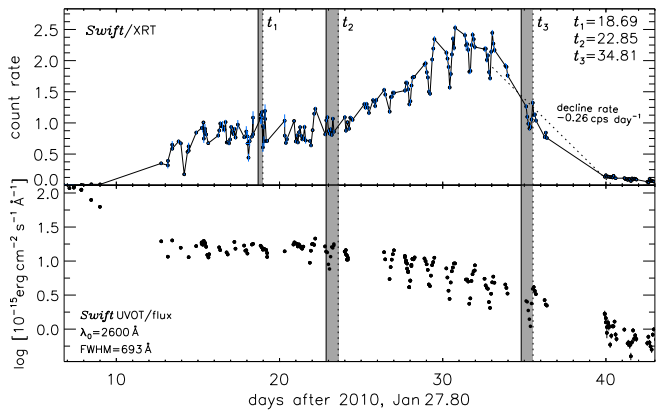}}
\caption{\label{swlc}\swift/XRT (top) and UVOT (bottom) light curves
of U\,Sco. The shaded regions mark the duration of continuous, deeper
\chandra\ ($t_1$) and \xmm\ ($t_2$ and $t_3$) observations.
}
\end{figure}

\swift\ X-ray and UV monitoring started within a day after discovery
(\citealt{atel2419}, but there was no
X-ray detection until 2010 Jan 31 (day 4). The \swift\ X-ray and UV light
curves shown in Fig.~\ref{swlc} will be discussed in more detail by
Pagnotta et al. (in preparation).
Until $\sim$day 8, the X-ray light curve was essentially constant with
weak, hard emission above 3\,keV \citep{atel2419}. On $\sim$day 12, the
\swift\ count rate was found to be 100 times higher, exhibiting a
supersoft spectrum \citep{atel2430} that resembles the class
of supersoft sources (SSS, \citealt{kahab,heuvel}). At the same
time, until $\sim$day 12, the UV flux dramatically decreased,
supporting the picture of a shrinking photospheric radius, accompanied
by a continuous increase in photospheric temperature. Between days
12 and 25, no changes in X-ray and UV brightness were seen, while
after day 25, another increase in X-rays with a simultaneous decrease
in UV emission level occurred.
From day 14.25 - 17.54 after discovery, two well covered,
and one partly covered, eclipses were seen in the UV light
curve by \cite{ATel2442} who also reported that, in X-rays,
no clear, sharp, deep or total eclipses were seen. The X-ray
light curve was highly variable with typically lower flux around
the time of eclipse. Around 2010 Mar 8 (day $t_0\,+\sim 40$\,days),
the brightness of
U\,Sco experienced a sharp drop in the X-ray/UV/Optical/IR
bands \citep{ATel2477}, indicating that nuclear burning had
turned off.

Reports of deeper and continuous \chandra\ and \xmm\
observations taken on
2010 February 14 and 19 ($t_1=t_0+18.7$ and $t_2=t_0+22.9$ days),
respectively were given by \cite{atel2451} and
\cite{atel2469}.
The times of these observations plus a
second \xmm\ observation taken on $t_3=t_0+34.9$ days
are marked by gray shaded areas in Fig.~\ref{swlc}, bracketing the
respective start and stop times. High-resolution X-ray grating
spectra consist of photospheric continuum emission plus broad
emission lines that varied greatly between the two observations
\citep{atel2469}, yielding a systematic increase in flux in lines
of higher ionization stage.

\section{X-ray/UV observations}

\begin{table*}[!ht]
\begin{flushleft}
\caption{\label{obslog}Observation log}
\begin{tabular}{llcll|llcllr}
\hline
\multicolumn{5}{l}{\bf ObsID 0650300201} & \multicolumn{5}{|l}{\bf ObsID 0561580301}\\
& Mode  & Start Time (UT) & Exp & counts & & Mode  & Start Time (UT) & Exp & \multicolumn{2}{c}{counts}\\
&& MM-DD@HH:MM:SS&(s) & s$^{-1}$& & & MM-DD@HH:MM:SS&(s)&\multicolumn{2}{c}{s$^{-1}$}\\
\hline
 MOS1 &   Large Window & Feb-19@15:41:19 & 63573 & 4.37$^a$ & MOS1  & Small Window &  Mar-03@14:33:51 & 62557 & \multicolumn{2}{c}{5.47$^a$}\\
 MOS2 &   Timing Window & Feb-19@15:41:30 & 63308 & 4.28$^a$ & MOS2  & Small Window &  Mar-03@14:33:51 & 62562 & \multicolumn{2}{c}{5.56$^a$} \\
 pn   &   Small Window & Feb-19@15:47:07 & 63365 & 25.90$^a$ & pn    & Small Window &  Mar-03@14:39:24 & 62364 & \multicolumn{2}{c}{28.45$^a$}\\
 RGS1 & Spectroscopy & Feb-19@15:40:20 & 63785 & 0.91$^b$ & RGS1  & Spectroscopy &  Mar-03@16:52:22 & 54419 & \multicolumn{2}{c}{1.15$^b$} \\
 RGS2 & Spectroscopy & Feb-19@15:40:25 & 63780 & 0.85$^b$ & RGS2  & Spectroscopy &  Mar-03@14:32:42 & 62779 & \multicolumn{2}{c}{1.06$^b$} \\
 OM   &   UV grism   & Feb-19@15:46:02 & 2000  & 11.1$^c$ & OM    & UVW1 &  Mar-03@14:38:19 & 2500 & 10.46$^d$ & 10.0$^e$\\
 OM   &   UV grism   & Feb-19@16:24:29 & 2000  & 11.7$^c$ & OM    & UVW1 &  Mar-03@15:25:06 & 2500 & 10.60$^d$ & 9.93$^e$\\
 OM   &   UV grism   & Feb-19@17:02:56 & 2000  & 11.6$^c$ & OM    & UVW1 &  Mar-03@16:11:53 & 2500 & 10.32$^d$ & 9.57$^e$\\
 OM   &   UV grism   & Feb-19@17:41:23 & 2000  & 10.0$^c$ & OM    & UVW1 &  Mar-03@16:58:40 & 2500 & 9.22$^d$ & 8.55$^e$\\
 OM   &   UV grism   & Feb-19@18:19:50 & 2000  & 8.81$^c$ & OM    & UVW1 &  Mar-03@17:45:27 & 2500 & 8.64$^d$ & 8.04$^e$\\
 OM   &   UV grism   & Feb-19@18:58:17 & 2000  & 7.27$^c$ & OM    & UVW1 &  Mar-03@18:32:14 & 2500 & 7.82$^d$ & 7.47$^e$\\
 OM   &   UV grism   & Feb-19@19:36:44 & 2000  & 5.64$^c$ & OM    & UVW1 &  Mar-03@19:19:01 & 2500 & 7.51$^d$ & 7.03$^e$\\
 OM   &   UV grism   & Feb-19@20:45:12 & 2000  & 6.28$^c$ & OM    & UVW1 &  Mar-03@20:05:48 & 2500 & 7.50$^d$ & 7.07$^e$\\
 OM   &   UV grism   & Feb-19@21:23:39 & 2000  & 8.77$^c$ & OM    & UVW1 &  Mar-03@20:52:35 & 2500 & 6.94$^d$ & 6.67$^e$\\
 OM   &   UV grism   & Feb-19@22:02:06 & 2000  & 9.75$^c$ & OM    & UVW1 &  Mar-03@21:39:22 & 2500 & 7.01$^d$ & 6.08$^e$\\
 OM   &   UV grism   & Feb-19@22:40:33 & 2000  & 10.9$^c$ & OM    & UVW1 &  Mar-03@22:26:09 & 2500 & 6.32$^d$ & 6.09$^e$\\
 OM   &   UV grism   & Feb-19@23:19:00 & 2000  & 12.3$^c$ & OM    & UVW1 &  Mar-03@23:12:56 & 2500 & 6.62$^d$ & 6.12$^e$\\
 OM   &   UV grism   & Feb-19@23:57:27 & 2000  & 12.3$^c$ & OM    & UVW1 &  Mar-03@23:59:43 & 2500 & 6.21$^d$ & 5.81$^e$\\
 OM   &   UV grism   & Feb-20@00:35:54 & 2000  & 12.5$^c$ & OM    & UVW1 &  Mar-04@00:46:30 & 2500 & 4.89$^d$ & 4.67$^e$\\
 OM   &   UV grism   & Feb-20@01:14:21 & 2000  & 12.4$^c$ & OM    & UVW1 &  Mar-04@01:33:17 & 2500 & 4.34$^d$ & 4.09$^e$\\
 OM   &   UV grism   & Feb-20@01:52:48 & 2000  & 13.3$^c$ & OM    & UVW1 &  Mar-04@02:20:04 & 2500 & 3.48$^d$ & 3.31$^e$\\
 OM   &   UV grism   & Feb-20@02:31:15 & 2000  & 15.1$^c$ & OM    & UVW1 &  Mar-04@03:06:51 & 2500 & 2.76$^d$ & 2.32$^e$\\
 OM   &   UV grism   & Feb-20@03:09:42 & 2000  & 12.5$^c$ & OM    & UVW1 &  Mar-04@03:53:38 & 2500 & 3.24$^d$ & 3.02$^e$\\
 OM   &   UV grism   & Feb-20@03:48:09 & 1980  & 12.1$^c$ & OM    & UVW1 &  Mar-04@04:40:25 & 2500 & 4.81$^d$ & 4.51$^e$\\
 OM   &   UV grism   & Feb-20@04:26:16 & 2000  & 13.3$^c$ & OM    & UVW1 &  Mar-04@05:27:12 & 3300 & 7.05$^d$ & 6.78$^e$\\
 OM   &   UV grism   & Feb-20@05:04:43 & 2000  & 13.4$^c$ & OM    & UVW1 &  Mar-04@06:27:19 & 3480 & 8.96$^d$ & 8.30$^e$\\
 OM   &   UV grism   & Feb-20@05:43:10 & 2000  & 13.4$^c$ &&&&&&\\
 OM   &   UV grism   & Feb-20@06:21:37 & 2000  & 13.0$^c$ &&&&&&\\
 OM   &   UV grism   & Feb-20@07:00:04 & 2000  & 12.2$^c$ &&&&&&\\
 OM   &   UV grism   & Feb-20@07:38:31 & 1940  & 13.3$^c$ &&&&&&\\
 OM   &   UV grism   & Feb-20@08:16:17 & 1800  & 13.0$^c$ &&&&&&\\
 OM   &   UV grism   & Feb-20@08:51:24 & 1700  & 12.8$^c$ &&&&&&\\
\hline
\end{tabular}

$^a$Average of background subtracted light curve from {\tt epiclccorr}\\
$^b$Average of background subtracted light curve from {\tt rgslccorr}\\
$^c$Integrated over dispersed grism spectra $\lambda=2200-3600$\,\AA\\
$^d$From {\tt omichain} using image data\\
$^e$Average of background subtracted light curve from {\tt omfchain} using fast window
\renewcommand{\arraystretch}{1}
\end{flushleft}
\end{table*}


 Two 16-hour \xmm\ observations were carried out with start times
2010 February 19.65 ($t_2=t_0+22.9$\,days) and 2010 March 3.7
($t_3=t_0+34.9$\,days). Simultaneous X-ray light curves (0.1-1\,keV),
X-ray spectra (5-38\,\AA\ range, 0.01\,\AA\ resolution), and ultraviolet (UV)
light curves and spectra (2200-3600\,\AA\ range, 15\,\AA\ resolution)
were extracted from the European Photon Imaging Camera (EPIC), the
Reflection Grating Spectrometer (RGS), and the Optical Monitor (OM),
respectively. For day 22.9, 27 UV grism spectra (approximately 2000
seconds exposure
time) were taken and were used to extract UV light curves by integrating
over three different wavelength intervals: the full range, 2200-3600\,\AA,
and the ranges 2800-3600\,\AA\ and 2200-2800\,\AA. In addition, we
extracted an
uncalibrated zero-order light curve from the non-dispersed photons
which represent the brightness over a broad wavelength band with
central wavelength $\sim 4000$\,\AA. Since the zero
order is not flux calibrated, count rates are only presented as an
indicator for relative brightness variations. For day 34.9, only
band-integrated fluxes are available but in high time resolution of
the OM fast mode.

Separate X-ray light curves were extracted in a soft (0.1-0.5\,keV)
and a hard (0.5-1\,keV) band, $S$ and $H$, and were used to compute
a spectral hardness light curve $HR=(H-S)/(H+S)$.

\begin{figure*}[!ht]
\resizebox{\hsize}{!}{\includegraphics{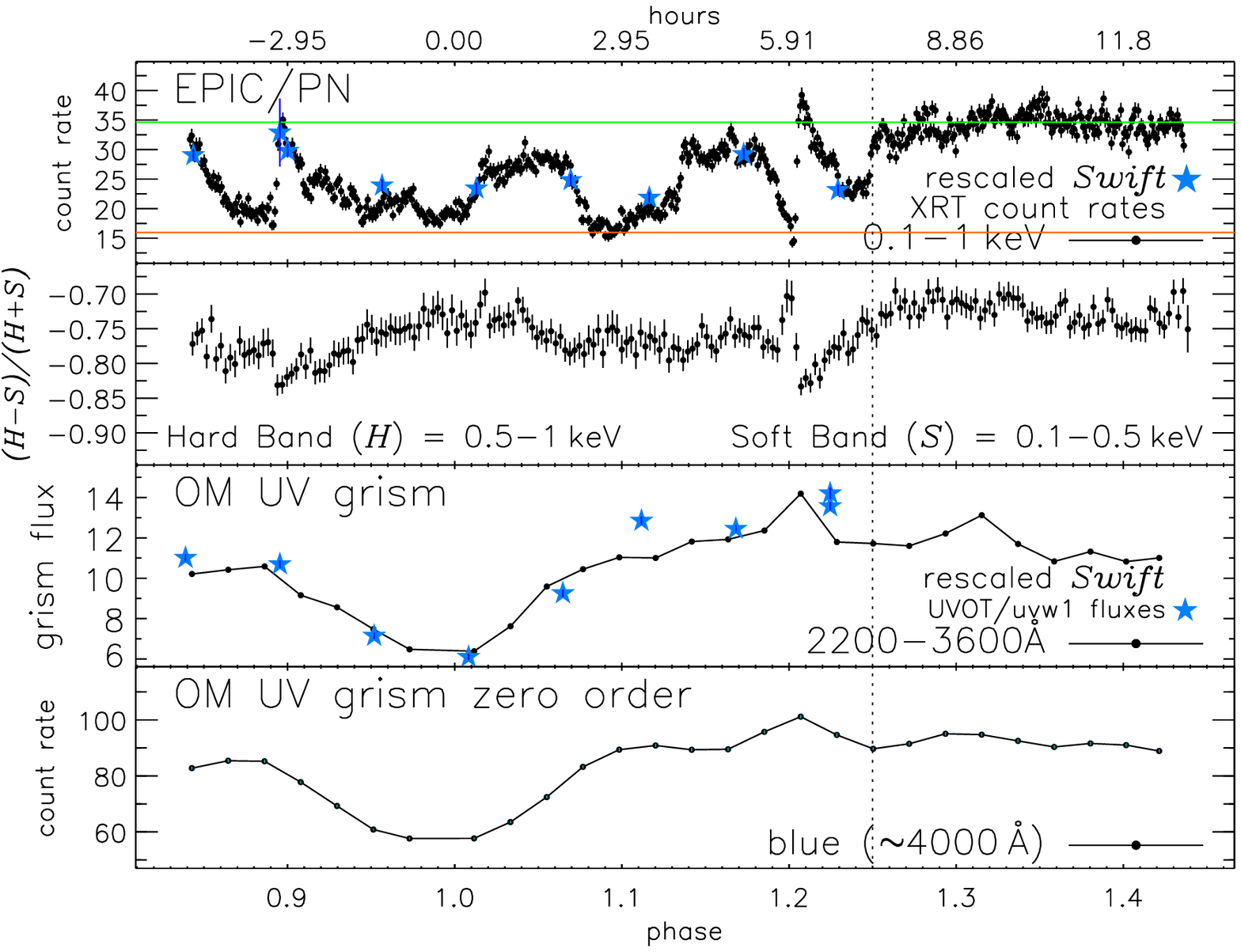}}
\caption{\label{lc1}Short-term evolution of X-ray and UV/optical
brightness and X-ray hardness (second panel) after $t_2=t_0+22.9$
days, with bandpasses given in vertical axes and legends.
Time units are phase, based on Eq.~\ref{ephem}. Phase 1.0 is when
the companion is between the white dwarf primary and the observer,
and the vertical dotted line indicates quadrature phase.
After phase 1.25, the companion is located behind the primary.
In the top horizontal axis, hours relative to phase 1.0 are given.
Simultaneous \swift\ data are added with blue star symbols, after
scaling by a factor 25.
UV grism fluxes are plotted in the third panel in units of
$10^{-15}$\,erg\,cm$^{-2}$\,s$^{-1}$\,\AA$^{-1}$.
}
\end{figure*}

\begin{figure}[!ht]
\resizebox{\hsize}{!}{\includegraphics{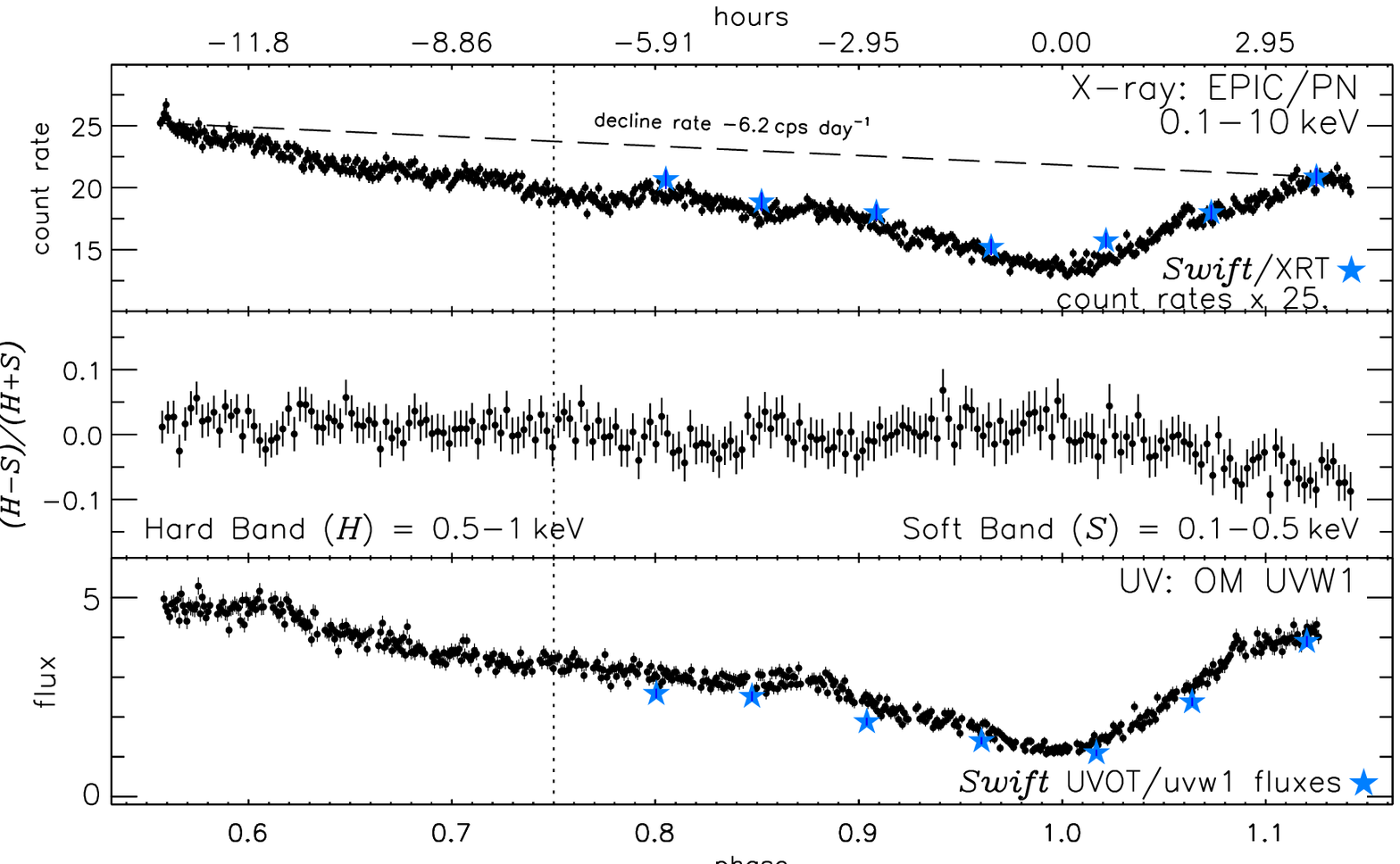}}
\caption{\label{lc2}Evolution of X-ray and UV brightness (top
and bottom) plus X-ray hardness (middle), taken on $t_3=t_0+34.9$
days, as in Fig.~\ref{lc1}. The dashed line in the top panel marks the
X-ray decline rate of 6.2\,cps day$^{-1}$ and is consistent, after
rescaling by the XRT-to-EPIC scaling factor of 25 (see caption of
Fig.~\ref{lc1}), with the long-term decline
rate in the \swift/XRT data (see Fig.~\ref{swlc}).
In the bottom panel, UV fluxes from \xmm/OM and \swift/UVOT
are shown in units of $10^{-15}$\,erg\,cm$^{-2}$\,s$^{-1}$\,\AA$^{-1}$.
In both light curves, a small step is seen at phase
$\sim 0.6$ and the light curves are asymmetric, yielding a sharper
egress than ingress. This could indicate an occultation
of the central source by an accretion stream that is located
ahead of the path of the companion.
}
\end{figure}

\section{Light curves}
\label{obslc}

In Fig.~\ref{lc1}, a time series of X-ray brightness, X-ray hardness,
UV flux, and zero-order (optical blue) count rate are shown as a
function of orbital phase, based on Eq.~\ref{ephem}. The
optical and UV light curves (bottom two panels, error bars are smaller
than the plot symbols) contain clean primary eclipses. The optical light
curve is smoother than the UV light curve. The X-ray light curve
(top panel) exhibits
high-amplitude variations between two brightness levels (horizontal green and
orange lines). The hardness ratio shown in the second panel is
less variable than the X-ray brightness. A broad, flat peak may be seen that is
centered around phase 1.0, indicating that softer emission from the
central source may be eclipsed. The event at phase 1.21 could
be a flare-like brightening, as anomalies are seen in all bands plus in the
spectral hardness light curve; for more discussion on the spectral
evolution, see Sect.~\ref{specevol}.
Possibly, also the sharp increase in X-rays plus reduction in spectral
hardness at phase 0.9 could indicate a flare, but no simultaneous
signatures in the UV/optical or hardness light curves support that.
Note that similar sudden brightness increases have been seen in other
novae, e.g., V1494\,Aql \citep{drake03}.

 After phase 1.25 (quadrature), the count rate
stabilized at the high-flux state for the remaining $\sim 20$\,ks
(=5.5 hours), $\sim 30$\% of the entire observation. This phenomenon
is worth closer inspection because after phase 1.25, the X-ray dark
secondary moves behind the X-ray bright photosphere which can then
no longer be eclipsed.
To investigate whether this could be a coincidence, we searched for
other X-ray observations taken outside of eclipse, but found no
similar occurrence in other data. The large majority of \swift\
observations (see Fig.~\ref{swlc}) were centered around eclipses, and no
continuous light curve outside eclipse can be assembled from these
data that could confirm or reject this conclusion. It is noteworthy that
\cite{ATel2442} report typically lower flux around the time of eclipse,
based on \swift/XRT data covering several eclipses. The \chandra\ light curve
extracted from the observation on $t_1=t_0+18.7$ days was
taken outside eclipse, and no such variations were reported by
\cite{atel2451}, but this observation was quite short
(15\,ks) and much earlier in the SSS phase.
Although coincidence cannot be completely ruled out, we consider
the fact that one third of the entire observation shows no dips
as support for the conclusion that the variations are related to the
binary orbit and therefore, that the occulted plasma resides inside the
binary orbit.

 Except for the two events at phases 0.9 and 1.21, we consider
the patterns of variability as flux decrements (dips) with
the undisturbed flux level being that observed after phase 1.25.
An inhomogeneous brightness distribution that is eclipsed by
the companion can securely be excluded, because during four minima
the flux level drops to 50\%. Each of four bright regions would
have to contribute 50\% to the total flux which is not possible.

 The similar flux levels during minima and maxima leads us to
the conclusion that the same source
is occulted multiple times by different occulters. Since during the
minima the flux is not zero, the central source can not contribute
more than $\sim 50$\% to the total X-ray emission, while the rest
must originate from further away, e.g., caused by scattering in the
outer regions of the ejecta
(see also Sects.~\ref{rgsspec} and \ref{specevol}).

 Within this picture, the totally different behavior in
the UV and optical can be explained by occulting plasma that
is transparent to UV, only leaving visible a single eclipse by
the companion. For example, neutral hydrogen is transparent
to UV/optical light but opaque to soft X-rays.
Alternatively, the X-ray emitting material that is likely more
concentrated to around the white dwarf suffers sharper occultations
by clumps while the larger optical emitting region will not have any
significant fraction covered.

 During the second \xmm\ observation on day 34.9, the
nova was already in the process of turning off, as is evident
from a rapid decline in X-ray and UV brightness observed by
\swift\ (Fig.~\ref{swlc} and \citealt{ATel2477}). This can also be
seen in the \xmm\ light curves in Fig.~\ref{lc2}, where the
X-ray and UV light curves are shown in the top and bottom panels,
respectively. \swift/XRT count rates, scaled by a factor 25
to compensate for different sensitivity, are also shown.
The X-ray decline rate is marked
by a dotted line with text above it. After
conversion for different sensitivities, this rate is consistent
with the long-term decline rate in the \swift/XRT light curve
that is marked in Fig.~\ref{swlc}.

The \swift/UVOT fluxes are consistent with the OM
fluxes (bottom panel). A small step at phase $\sim 0.62$ can
not be explained by an eclipse by the companion, but it
could be an eclipse by an accretion stream (see \S\ref{disc}).

A clean eclipse can be seen in all light curves centered around
phase 1. The hardness light curve shown in the middle panel does
not change with the eclipse. Towards the end of the observation,
the source becomes slightly softer, possibly related to the
decline.

 The dips have disappeared from the X-ray light curve, and must
therefore be understood as a transient phenomenon during the
early SSS phase. This behavior appears similar to RS\,Oph and
other novae that have been monitored with \swift, showing an
early variability phase. The \xmm\ data of U\,Sco give
valuable insights into the processes that may explain such
an early variability phase in more general terms.

\subsection{Light curve model}
\label{lcmodel}

As a qualitative test of the interpretation of occultations of
the central source by dense absorbers, we have constructed a
geometrical model which consists of two sources of light and four
occulters. On a $350\times 350$ brightness pixel map, two sources
of light are defined, a non-variable, spherical central source with
a radius $R_{\rm primary}$ plus additional constant emission that
is not subject to occultations. Light
originating from the central source is removed by the secondary
star or by three clumps of gas that is optically thick to X-rays.
All occulters are assumed to be spherical. In this way, a synthetic
light curve can be generated by calculating for each phase bin the
difference between the sources of light and the light originating
from those pixels located behind the absorbers as seen from Earth.

 The inclination angle and radii of the orbit and companion
are well known and are fixed parameters. Unknown parameters
are the sizes of the central source and the absorbers, the
location of the absorbers, and their opacity (see next
paragraph). Values of these parameters were obtained by a
combination of manual and automatic iteration with the aim
to obtain good agreement between the synthetic and the
observed light curves.
 Many different configurations are possible such as a
non-spherical geometry of the source or the absorbers as well
as intrinsic variability. Under these circumstances, we do not
believe that beyond a qualitative feasibility test, reliable
quantitative conclusions can be derived.

 The opacity of the absorbers is parameterized by the
column density of neutral atomic hydrogen, $N_{\rm H}$.
Depending on this value, a certain fraction of soft X-ray
emission that originates from behind a given absorber
can penetrate, yielding full transparency for $N_{\rm H}=0$.
The fraction of light that can pass through the
absorber, $T(w_i)$, depends on wavelength $w_i$:
$T(w_i)=e^{-N_{\rm H}\times \sigma(w_i)}$, where $\sigma(w_i)$
is the cross section for a given wavelength bin $w_i$
and was computed using the {\tt bamabs} tool which
is part of the PintofAle package \citep{pintofale}.
The fraction of absorbed light, integrated over the entire
spectrum, $sp(w_i)$, is
$F=1-\sum_i \{T(w_i)\times sp(w_i)\}/\sum_i sp(w_i)$.

\begin{figure}[!ht]
\resizebox{\hsize}{!}{\includegraphics{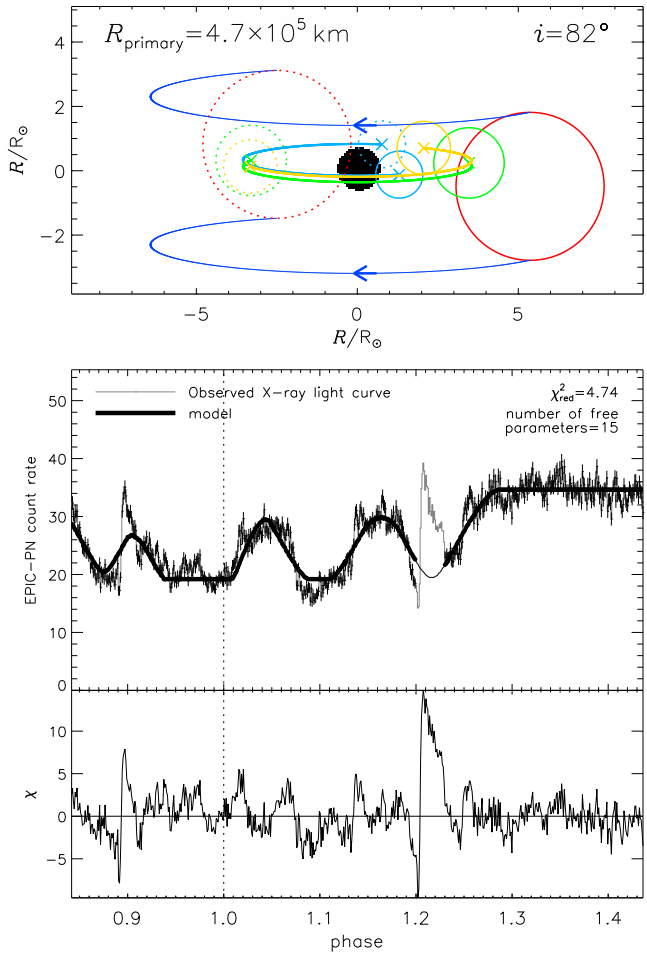}}
\caption{\label{emap22}Qualitative geometrical model to the
X-ray light curve taken on day 22.9. The top panel gives a spatial
representation of the model as seen from Earth (inclination angle
given in top right legend) and is to scale with
units given in the axis labels. The location and size of the central
source is marked by the black filled circle, centered at the origin
of the coordinate system and the assumed radius given in the top left
legend. The central source is eclipsed by the companion (red circle)
and by three co-rotating absorbers (blue, green, and orange circles).
The sizes and positions of the circles are to scale with the solid
and dotted styles denoting the respective positions at the start and
the end of the observation.
Eclipses by any of these occulters lead to temporary decreases in total
light that, in the absence of eclipses, is assumed to be constant.
Additional emission is added that remains constant, regardless
of binary phase.
The resulting model light curve is shown in comparison to the data
in the middle panel. The bottom panel shows the difference between
data and model, relative to the measurement errors,
$\chi=$(data-model)/$\Delta$.
Except for two events at phases 0.9 and 1.2, the residuals between
model and data are of the same order as the variations outside
eclipse, past phase 1.25.
}
\end{figure}

 In Fig.~\ref{emap22}, a graphical illustration of a best-fit
model to the X-ray light curve from day 22.9 is shown in the top.
Below, the resulting light curve in comparison to the
observed X-ray light curve and the residuals relative to the
measurement errors ($\chi$) are shown. Except for the
flare-like event at phase 1.2 and
the sharp rise at phase 0.9, all residuals are consistent with
the degree of variability after phase 1.25 and can be considered
as not related to occultations.
Further refinements of this model will not increase
the precision of the conclusion.

\begin{figure}[!ht]
\resizebox{\hsize}{!}{\includegraphics{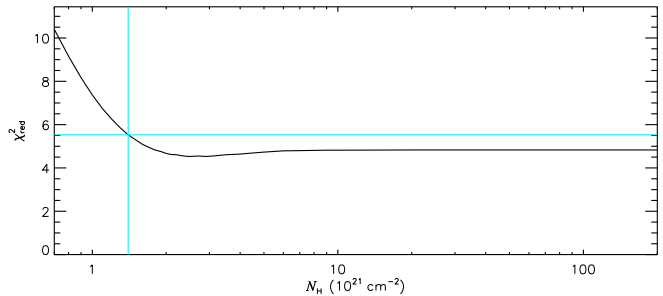}}
\caption{\label{nh_chi}Determination of column density of
absorbers in light curve model. While the light model yields
poor resemblance with observations for low values of $N_{\rm H}$,
any high value of $N_{\rm H}$ is consistent with the data.
}
\end{figure}

While the degree of reproduction is not great,
$\chi^2_{\rm red}=4.7$, we attribute the deficiencies to additional
random intrinsic variability that can not be parameterized.
The parameters of this model are, for the central source,
a radius of $R_{\rm primary}=0.67$\,R$_\odot$ (17\% of the Roche
Lobe) and a brightness level
of 16.6 counts per second. A power-law surface brightness distribution
with a power law index $\alpha$ was assumed, but the index
came out very small with $\alpha=-5.8\times 10^{-5}$. We
experimented with limb darkening or -brightening but found no
better reproduction of the data. The level of additional constant
emission was 17.7 counts per second. The nature of the additional
non-variable emission is discussed in the context of the spectra in
Sect.~\ref{rgsspec} and \ref{specevol}. The absorbers are assumed to
orbit at a radius of 3.55\,R$_\odot$, having a size of
0.73, 1.1, and 0.83\,R$_\odot$, respectively. For the column
density in the absorbers, a lower limit of
$N_{\rm H}>1.4\times 10^{21}$\,cm$^{-2}$ is determined. This is
illustrated in Fig.~\ref{nh_chi} from which it can be deduced that
$N_{\rm H}$ is unconstrained at high values. We conclude that
the absorbers are completely opaque to the X-ray spectrum of
U\,Sco, and the lower limit represents the value above which
only the constant additional emission remains.
This can be compared to predictions from the expected
density of the accretion disk material and the path length
through it. The volume of a flared disk rim is of the
order of $10^{34}$ cm$^3$, e.g., \cite{drake10}. For a mass
loss rate of $\sim 10^{-7}$\,M$_\odot$\, yr$^{-1}$ (e.g.,
\citealt{hachisu_usco}), and a build-up time of $\sim 5$ days,
the accretion disk rim density is of order $3\times10^{14}$\,cm$^{-3}$.
For a path length through the gas of $\sim 0.1$ R$_\odot$, the column
density is of order 10$^{24}$\,cm$^{-2}$. While this is higher than
the number derived from the eclipse model, a fraction of all
hydrogen can be ionized, requiring a higher column density for
the same opacity, or the build-up time may be shorter.
Moreover, the disk may not have fully reformed at the time
of observation, and during the reformation process, the
density should be lower.

The vertical dotted line in the bottom two panels of
Fig.~\ref{emap22} indicates phase 1.0, where the center of
eclipse by the companion should be. A dip of the expected width
is present, however, it is clearly shifted. For better reproduction
of the data, we had to include a phase shift parameter with a value
of -0.026 in the model. Such an artificial shift could be avoided if
a non spherical symmetry of the central X-ray source is allowed. 
For example, we tested a model with a crescent shaped source and were
able to reproduce the second dip without a phase shift.
There is, however, no physical explanation for a crescent
shaped central source.

\begin{figure}[!ht]
\resizebox{\hsize}{!}{\includegraphics{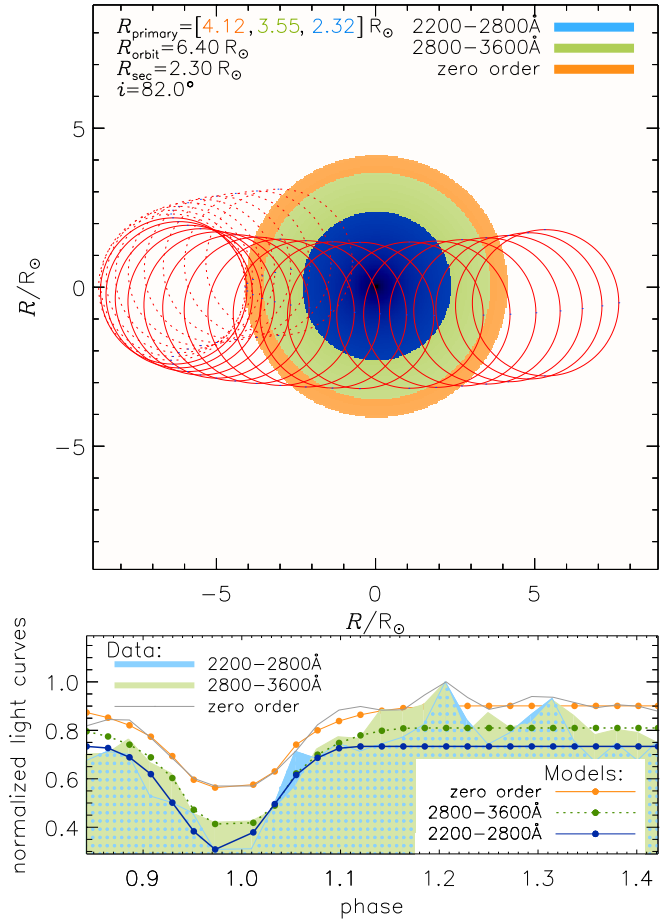}}
\caption{\label{emap_uv}Geometrical light curve model applied
to three UV/optical light curves extracted from the non-dispersed
zero grism order (optical, orange) and two UV light curves (green
and blue), extracted from
different wavelength ranges as denoted in the upper right legend.
The central source was assumed spherical with a power-law
brightness distribution and the resulting best-fit radii for each
wavelength band given in the top left legend. The three emitting
regions are illustrated as filled circles. The location and size
of the companion is plotted as a red circle at the beginning of
each of the 27 exposures.
In the {\bf bottom panel}, the three observed light curves (shades)
and the best-fit models (points) are shown in normalized units.
}
\end{figure}

\begin{figure}[!ht]
\resizebox{\hsize}{!}{\includegraphics{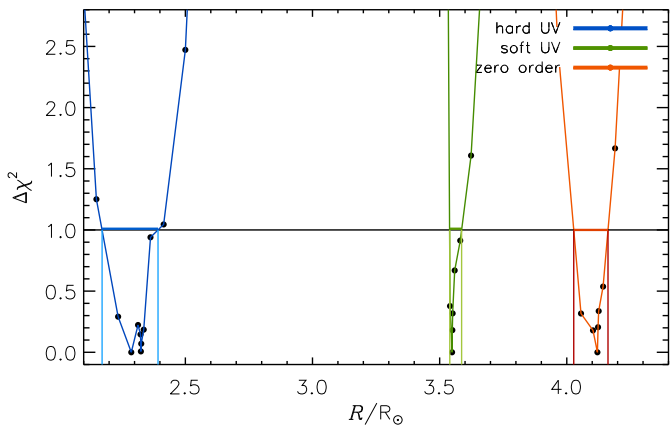}}
\caption{\label{chi}Estimate of uncertainties in the
primary radius obtained from three UV/optical light curves,
using the color scheme defined in Fig.~\ref{emap_uv} (see
legend).
}
\end{figure}

 The same procedure was applied to the 27 OM grism exposures of
day 22.9. We constructed two separate UV light curves by integrating
each grism spectrum over the two wavelength bands 2200-2800\,\AA\ and
2800-3600\,\AA. Combined with the
light curve from the non-dispersed photons, we have three light
curves representing different spectral colors, all containing
clear eclipses. In Fig.~\ref{emap_uv}, the results are illustrated,
where the red circles represent the location and relative size
of the companion during each of the 27 exposures and the shaded
circles in the center represent the primary with the best-fit
radii for each band, following the color scheme given in the
top right legend. The best-fit radii are given in the top left
legend. The short-wavelength UV band yields the smallest radius,
while the longest-wavelength optical band (zero order) yields
the largest radius, close to the inner Lagrangian point. The
significance of the light curve fit is illustrated
in Fig.~\ref{chi}, where a range of primary radii is plotted against
changes in $\chi^2$ relative to the best fit for each wavelength band.
The boundaries around the 1-$\sigma$ uncertainties are marked by the
vertical colored lines, based on an increase of $\chi^2$ by 1
(horizontal black line). The best-fit light curve models
are shown in comparison to the data (in normalized units) in the
bottom panel of Fig.~\ref{emap_uv}. While the models (points
connected by lines, see bottom right legend) give no perfect
reproduction of the
data (shaded areas and gray line, see top left legend),
the width of the respective eclipses are well reproduced,
yielding reliable radius estimates.

 The trend of decreasing radius with decreasing wavelength
appears like a temperature gradient. Hotter plasma,
producing more short-wavelength emission, resides closer to
the center than cooler plasma.

\section{Spectra}

 The X-ray spectra from two types of spectrometers are
discussed in \S\ref{xrayspec}. High-resolution soft grating
spectra and low-resolution hard CCD spectra are available for
both observations. In addition, 27 UV grism spectra have been
taken during the first observations that are discussed in
\S\ref{uvspec}.

\subsection{The X-ray spectra}
\label{xrayspec}

 \xmm\ carries two types of X-ray spectrometers that give information
from different energy bands, yielding different sensitivity and
spectral resolution. The EPIC
consists of two types of CCD detectors, the MOS (Metal Oxide
Semi-conductor; \citealt{epic_mos}) and the pn \citep{epic_pn}.
Two identical MOS cameras, MOS1 and MOS2, are in operation behind
two separate mirrors, and they can be operated in different modes.
They are sensitive between 0.1-10\,keV with an effective area
ranging between $50-500$\,cm$^2$ and an
energy resolution of roughly 50\,eV. The pn has a lower
resolution of 100\,eV but a higher sensitivity of
$200-1300$\,cm$^2$ from $0.1-10$\,keV because of the
larger collecting area of an extra set of mirrors.

The dispersive spectrometers RGS1 and RGS2
\citep{rgs} share the light with the same mirrors as the MOS
cameras. They are both sensitive between $6-38$\,\AA\ ($0.3-2$\,keV)
with an
effective area between $20-60$\,cm$^2$ and a wavelength
resolution of 0.05\,\AA. The dispersed photons are recorded
by MOS type CCDs. The intrinsic energy resolution of the
detectors is used to separate higher dispersion orders from
the first dispersion order spectrum. While the sensitivity
of the RGS is much lower than the EPIC, the superior
resolving power of $\Delta E/E \sim 100-600$ compared to
$\Delta E/E<1$ for EPIC allows individual line transitions
to be identified.
For example, only in the RGS spectrum can an
absorption line spectrum be distinguished from an emission
line spectrum, provided the source is bright enough or the
exposure time is sufficiently long for enough signal to noise.
SSS spectra are extremely bright and are therefore ideally
suited for studies with the RGS. The EPIC spectra are
superior above $\sim 2$\,keV and are use to study the hard
Wien tail and to search for
harder emission that may be related to shocks in the
surrounding medium, or accretion.

 Separate spectra for different time intervals can be extracted by application
of filters on photon arrival times. We have extracted the RGS and
EPIC spectra from the entire observations and a series
of spectra from adjacent shorter time intervals in order to probe
for spectral time evolution. Times were converted to orbital phase using
Eq.~\ref{ephem}.

\subsubsection{High-resolution RGS spectra}
\label{rgsspec}

\begin{figure}[!ht]
\resizebox{\hsize}{!}{\includegraphics{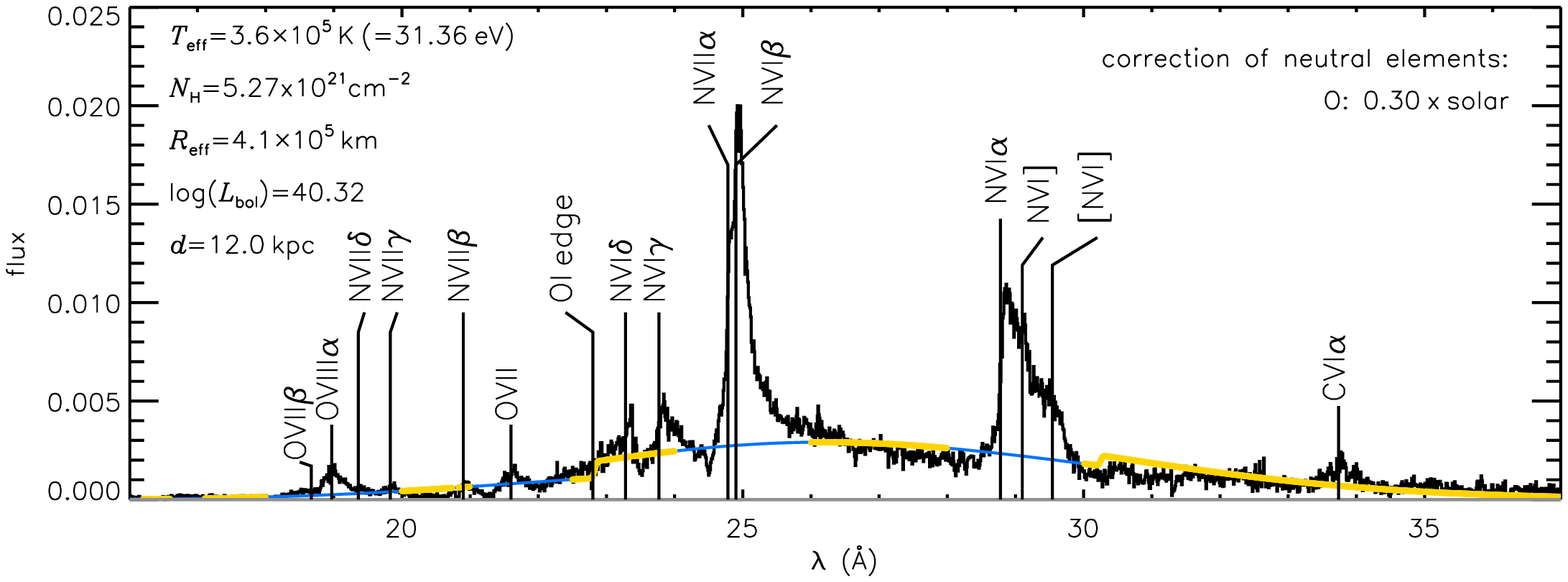}}

\resizebox{\hsize}{!}{\includegraphics{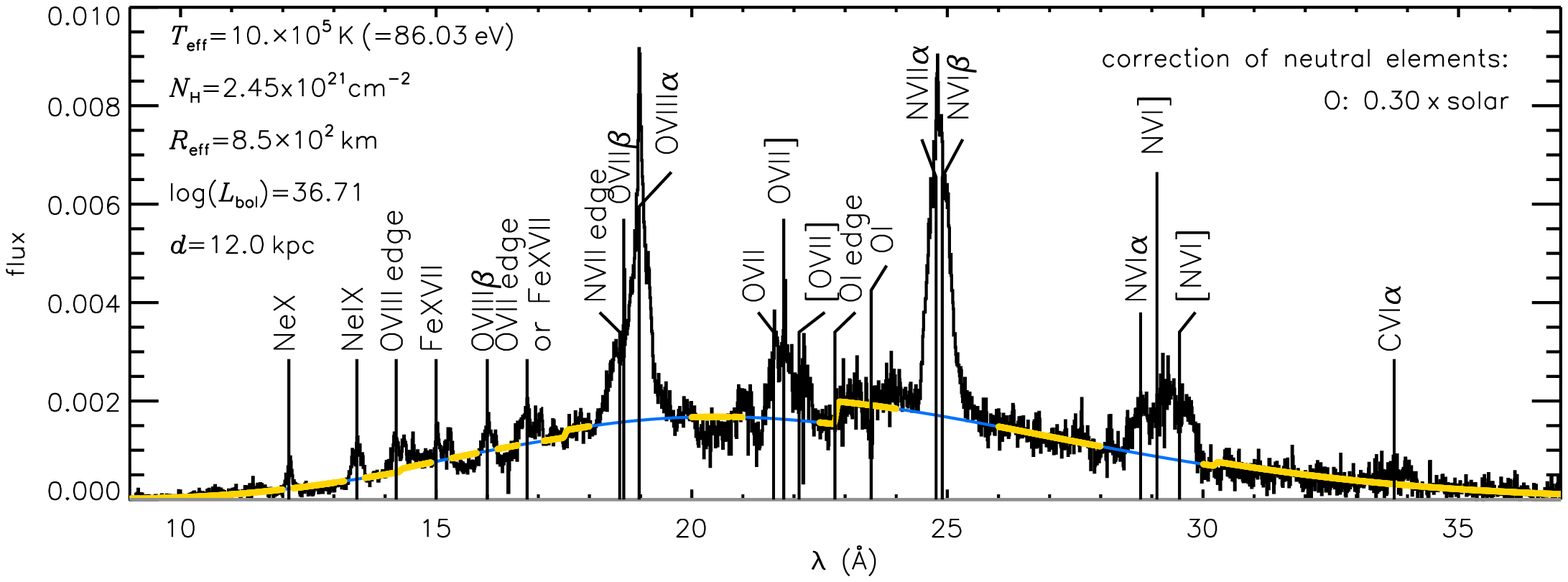}}
\caption{\label{bbfits}RGS spectra taken on day 22.9 (top) and 34.9
(bottom). Flux units are ph\,cm$^{-2}$\,s$^{-1}$\,\AA$^{-1}$. A
blackbody curve is included as blue/yellow line, where those
spectral bins that were used for fitting are yellow.
The best-fit parameter values are given in the left legends
but the main purpose of the blackbody fitting is to show that the
continuum component is of atmospheric origin.
The photospheric temperature has greatly increased from day 22.9
to day 34.9, evidenced by a clear shift of the Wien tail to
shorter wavelengths. Additional strong emission lines are marked
with labels. Lines of higher ionization stages are seen
on day 34.9, yielding additional evidence for a temperature increase.
}
\end{figure}

 The merged RGS1+RGS2 spectra from both observations are shown in
photon flux units in
Fig.~\ref{bbfits}. The spectra shown here have been produced with
the SAS command {\tt rgsfluxer}. The spectral range of the RGS
covers the peak of the absorbed photospheric
emission from the optically thick ejecta. A combination of
continuum emission and strong emission lines can be identified.

As a parameterization of the continuum component, we fitted a
blackbody model to those spectral bins that are not dominated
by strong emission lines. The resulting models are marked by the
yellow/blue curves in Fig.~\ref{bbfits} for each observation (see
caption).
 The resulting parameters effective temperature, interstellar
neutral hydrogen column density, and effective radius (derived from
dilution factor and assumed distance 12\,kpc) are given in the
left legends.
The bolometric luminosity, derived from these values applying
the Stefan-Boltzmann law, are also given but an unrealistically
high value is derived for day 22.9, exceeding the Eddington
luminosity by two orders of magnitude. This is a well-known
deficiency of blackbody fitting to SSS spectra
\citep[e.g.,][]{krautt96}, and we therefore caution that these
values can not be trusted to reflect the true physical conditions,
and they are only given for reproducibility. Appropriate models
have to account for radiation transport in non-local thermal
equilibrium (NLTE), plus
the expansion of the ejecta has to be taken into account
\citep[e.g.,][]{ness09,vanrossumness09}. These are currently
unsolved problems, and their discussion is beyond the scope of this
paper. Nevertheless, the reproduction of the continuum component
by a blackbody model is sufficiently good to allow the conclusion
that the continuum originates from the photosphere around the white
dwarf. The absence of any absorption lines could indicate that
we are not directly seeing the atmospheric emission. Gray scattering
in a plasma of fast electrons (Thompson scattering) could explain
this observation, as narrow atmospheric features would be smeared out.

 While the photospheric temperature can not be accurately
determined from blackbody fits, the location of the Wien tail
clearly indicates that the temperature must be well above
$10^5$\,K. From this lower limit of the temperature we can
conclude that the radial extent must be significantly smaller than
the radius of the Roche Lobe (4.1\,R$_\odot$). Otherwise, the bolometric
luminosity would be four orders of magnitude above the Eddington
limit which is physically unrealistic.\\

 The emission lines originate from highly
ionized nitrogen, oxygen and carbon. For N\,{\sc vii} and N\,{\sc vi},
the H-like and He-like 1s-$n$p series can be identified up to
$\delta$ (1s-5p). In the second observation, additional Ne\,{\sc ix}
and Ne\,{\sc x} lines can be identified.

In order to escape, the emission lines must originate in a reasonably
optically thin plasma, which can reside at any distance, far above the
photosphere. On the other hand, the strongest lines are seen at
wavelengths where the continuum is strongest, and they are thus
likely photoexcited, requiring that they come from hot plasma
residing close to the ionizing source.

 This type of X-ray spectrum, in both observations, is not
typical of novae such as V4743\,Sgr, RS\,Oph, or V2491\,Cyg
in which prominent deep photospheric absorption lines were seen by
\cite{v4743,ness_rsoph,ness_v2491}. It resembles more
that of the eclipsing supersoft source Cal\,87 which
\cite{ebisawa10} explain as a combination of Thompson scattered
continuum plus emission lines from resonant line scattering
as viewed at high inclination angle (see also \citealt{greiner_cal87}).
The photosphere around the white dwarf cannot be seen directly
as it is blocked by the accretion disk. A similar situation may be
occurring in U\,Sco and perhaps even in all high-inclination systems
as was already suggestive from UV observations presented by
\cite{dous91} and \cite{vitello93}. Thompson scattering
in the electron-rich environment would preserve the spectral
color, but a non-zero velocity distribution of electrons would
smear out photospheric absorption lines which can explain the
absence of any clear indication of absorption lines in the RGS
spectra of U\,Sco.

In addition to Thompson scattering, resonant line scattering may occur,
selectively amplifying or diminishing resonance emission lines, depending
on the geometry of the scattering plasma. For example, the geometry of
an accretion disk allows more photons from resonance transitions to be
scattered into the line of sight than out of the line of sight, and
resonance lines can appear amplified when compared to forbidden lines.
This is best seen in the He-like N\,{\sc vi} triplet at 28-29\,\AA,
where the resonance line is stronger than the intercombination line.
For comparison, in the
high-resolution X-ray spectra of low-density stellar coronal plasma,
the forbidden line is usually almost equally as strong
as the resonance line and stronger than the intercombination line
\citep{ness_cap}. While the low ratio of
forbidden to intercombination can be interpreted as high
densities \citep{gj69}, the same low ratio can also be caused
by strong UV radiation fields, if the emitting plasma resides
close to the photosphere. Inspection of dip and non-dip spectra
presented in Sect.~\ref{specevol} confirms this explanation.

\subsubsection{EPIC spectra}
\label{epicspec}
 
\begin{figure}[!ht]
\resizebox{\hsize}{!}{\includegraphics{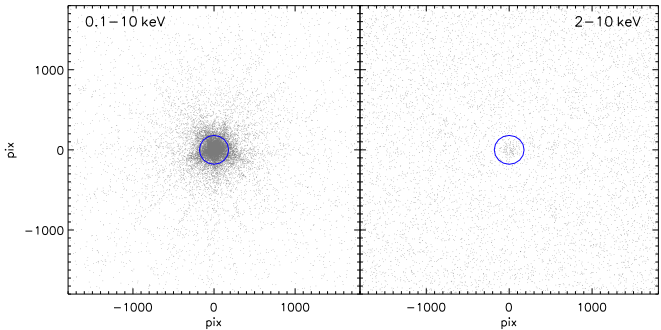}}
\caption{\label{photplot}EPIC/MOS1 observation taken on
day 22.9. In the left panel, the pixel positions of all events between
0.1-10\,keV are shown relative to the peak of the point spread 
function. In the right panel, only photons between 2 and 10 keV 
are shown. A concentration of counts towards the source position
can be seen, however, the inner 180 pixels (marked by the blue
circle around the center) are affected by pile up (see text).
}
\end{figure} 

\begin{figure}[!ht]
\resizebox{\hsize}{!}{\includegraphics{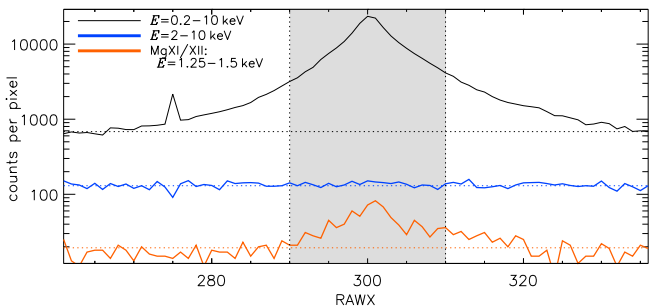}}
\caption{\label{mos2contour}EPIC/MOS2 timing mode observation taken on
day 22.9. In this mode, all photons along the
detector y axis (RAWY) are accumulated into a single number of counts
for each detector x-pixel (=RAWX) coordinate. The source spectrum was
extracted from between RAWX values 290-310 (marked by gray
shaded area), encircling 74\% of all counts.
Plotted is the number of counts per pixel, integrated over three
different energy ranges as indicated in the legend. Above
2\,keV, no significant detection is found. The energy range
1.25-1.5\,keV contains the H-like and He-like Mg lines.
}
\end{figure}

 EPIC spectra yield the best information about the high-energy part of
the spectrum, above 2\,keV, where the RGS is not sensitive. \swift\
detected hard emission early in the evolution \citep{atel2419}, and
it is of interest
to use the deeper \xmm\ EPIC data to check whether it was still
present on days 22.9 and 34.9.

 In Fig.~\ref{photplot}, the photon image of MOS1 taken on day 22.9
is shown with all events in the left panel and only those with
registered energies above 2\,keV in the right panel. While in the
right panel, a clear source seems to be present, this can not be
interpreted as the detection of hard emission. The blue circle
marks the radius within which significant pile up is present, which
was determined using the SAS tool {\tt epatplot}. All photons in the
right panel are within this circle, indicating that they are
contaminated by pile up. We encountered the same problem in the pn
observation and in all EPIC observations of the second observation.

 More sensitive to faint hard emission are the MOS2 data in the
first observation which was operated in timing mode (see
Table~\ref{obslog}), yielding a shorter readout time. We determined,
with {\tt epatplot}, that pile up was not a problem. In
Fig.~\ref{mos2contour}, the brightness profiles as a function of
the detector coordinates (RAWX) are shown for three different energy
ranges as indicated in the legend. The full-range profile clearly
indicates the location of the source in RAWX coordinates, and above
2\,keV, nothing is seen at the same position. In the narrow range
1.25-1.5\,keV, containing the H-like and He-like Mg lines, a clear
detection is found, indicating that some harder emission above the
SSS component is present. A source spectrum can be extracted from a range
in RAWX of 290-310 pixels, which is marked by the gray shaded area
in Fig.~\ref{mos2contour}. No other sources are located along the
y axis that could contaminate the source region. For our
further considerations, we therefore only use the MOS2 spectrum
for the first observation while for the second observation, we
use the most sensitive pn data, after correction for pile up.

\begin{figure}[!ht]
\resizebox{\hsize}{!}{\includegraphics{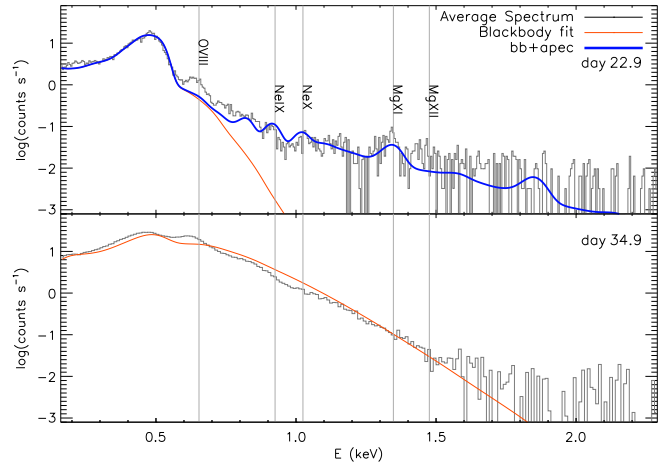}}
\caption{\label{epicspectra}EPIC spectra taken on days 22.9 (top, MOS2
detector) and 34.9 (bottom, pn detector).
The energies of strong H-like and He-like transitions are marked,
and the red lines indicate blackbody models with the same parameters
as used in Fig.~\ref{bbfits}. For day 22.9, a model
with an additional optically thin collisional component (APEC) is plotted
with a blue thick line.
}
\end{figure}

 The EPIC spectra for days 22.9 and 34.9 are shown in Fig.~\ref{epicspectra}.
The energies of some line transitions that have been seen in other novae
are marked with vertical lines with labels.

 The SSS continuum appears featureless at EPIC energy
resolution. Blackbody models with the same parameters as used in
Fig.~\ref{bbfits} are overplotted with the red curves. Obviously,
the EPIC spectra are not suitable for studies of the soft part below
2\,keV, even
if they are not piled up. However, in the first observation,
there is significant excess emission above the blackbody curve
at energies $\stackrel{>}{_\sim} 0.7$\,keV, indicating the presence
of an additional plasma component. This excess can be modeled with an
optically thin thermal APEC model \citep{smith01} as indicated by the
blue thick curve in Fig.~\ref{epicspectra}.
The APEC model component assumes collisional equilibrium for ionization
and excitation processes which is observable as bremsstrahlung continuum plus
emission lines. Optically thin emission has been seen in various novae
before, see, e.g., \cite{lloyd92,krautt96,mukai01,Orio2001}, and can be
caused by shocks with pre-existing surrounding material, e.g., from
previous outbursts, or within inhomogeneous ejecta.
The electron temperature obtained from fitting an
APEC model plus a blackbody component with fixed parameters to the
spectrum is k$T=0.29\,\pm\,0.03$\,keV. This temperature corresponds to an
average random electron velocity of $10^9$\,cm\,s$^{-1}$ when assuming a
Maxwellian velocity distribution, which is approximately twice the
expansion velocity. The H-like and He-like Ne
and Mg 1s-2p lines are well reproduced by the APEC model component.
The O\,{\sc viii} line at $\sim 0.66$\,keV (18.97\,\AA) is not
reproduced by the combined bb+apec model and is more likely
photoionized by the SSS continuum. In the second observation
the collisional component has faded significantly.


 Above 2\,keV, we found no significant continuum emission nor
Fe fluorescent lines and lines of Fe\,{\sc xxv} at 6.4 and 6.7\,keV,
respectively.
 A more detailed discussion of the evolution of the hard emission
will be instructive in the context of all X-ray observations,
including the early \swift\ and \suzaku\ data.

\subsubsection{Spectral Evolution}
\label{specevol}

{\bf A) Long term evolution}

 A comparison of the spectra from the two observations yields the
spectral evolution on time scales of days to weeks. A trend of increasing
temperature has been reported by \cite{atel2469} based on increasing
strengths in high ionization lines from day 18.7 to day 22.9.
A gradual increase of the effective temperature is expected as the
photospheric radius continues to shrink during the SSS phase.
From the spectra shown in Figs.~\ref{bbfits} and
\ref{epicspectra}, this trend can be seen to have continued through
day 34.9, evidenced by a clear shift of the Wien tail of the continuum
to higher shorter wavelength and the stronger emission lines
belonging to high-ionization stages (Fig.~\ref{bbfits}).
Only the high-ionization He/H-like Mg lines are stronger on
day 22.9 than on day 34.9 (Fig.~\ref{smap_epic}). Since they
are not formed on top of any continuum emission, they can not be
photoexcited. The more likely origin of the Mg and Ne lines is
a collisional plasma, and this interpretation is supported by the
APEC model.\\

\begin{figure*}[!ht]
\resizebox{\hsize}{!}{\includegraphics{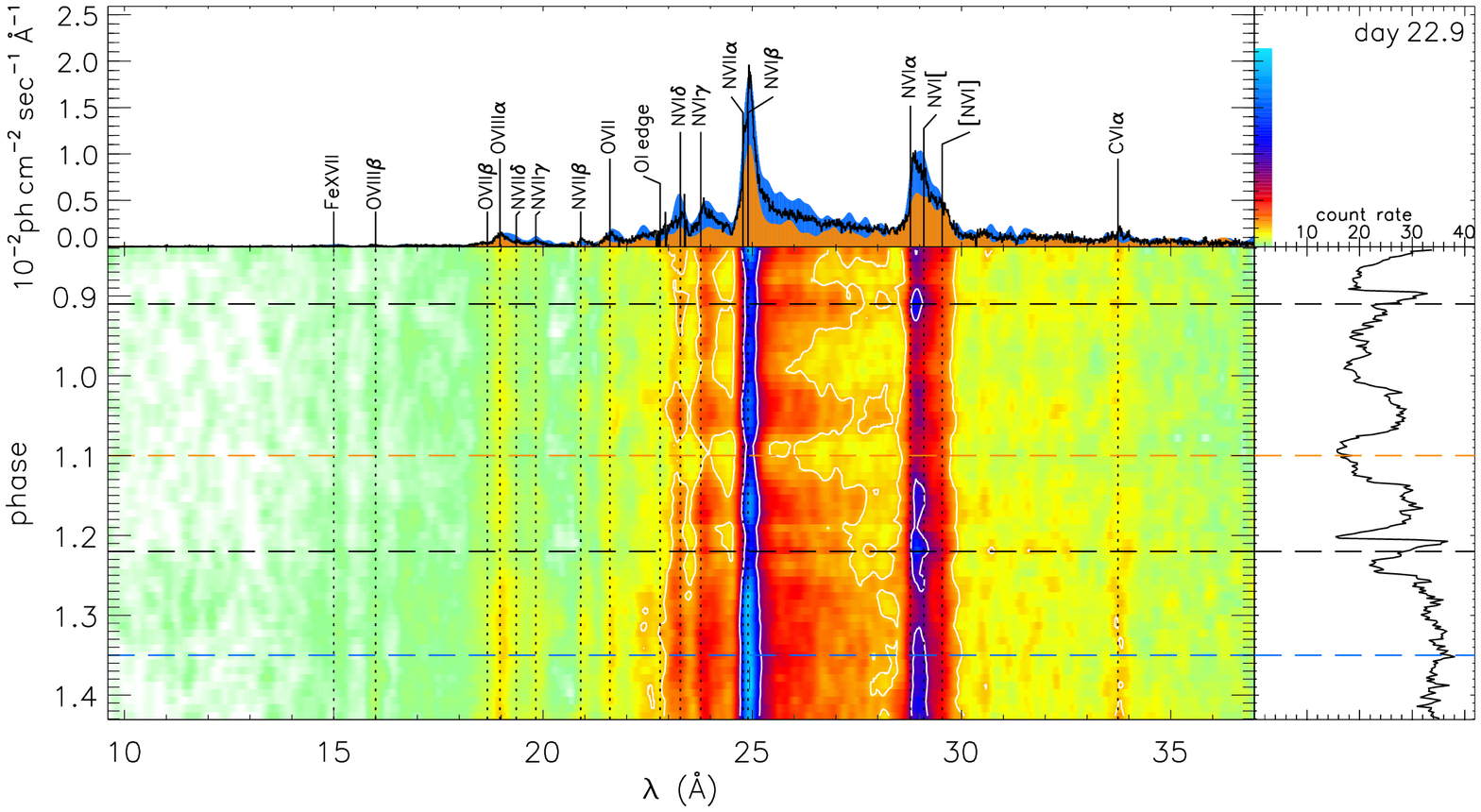}}

\resizebox{\hsize}{!}{\includegraphics{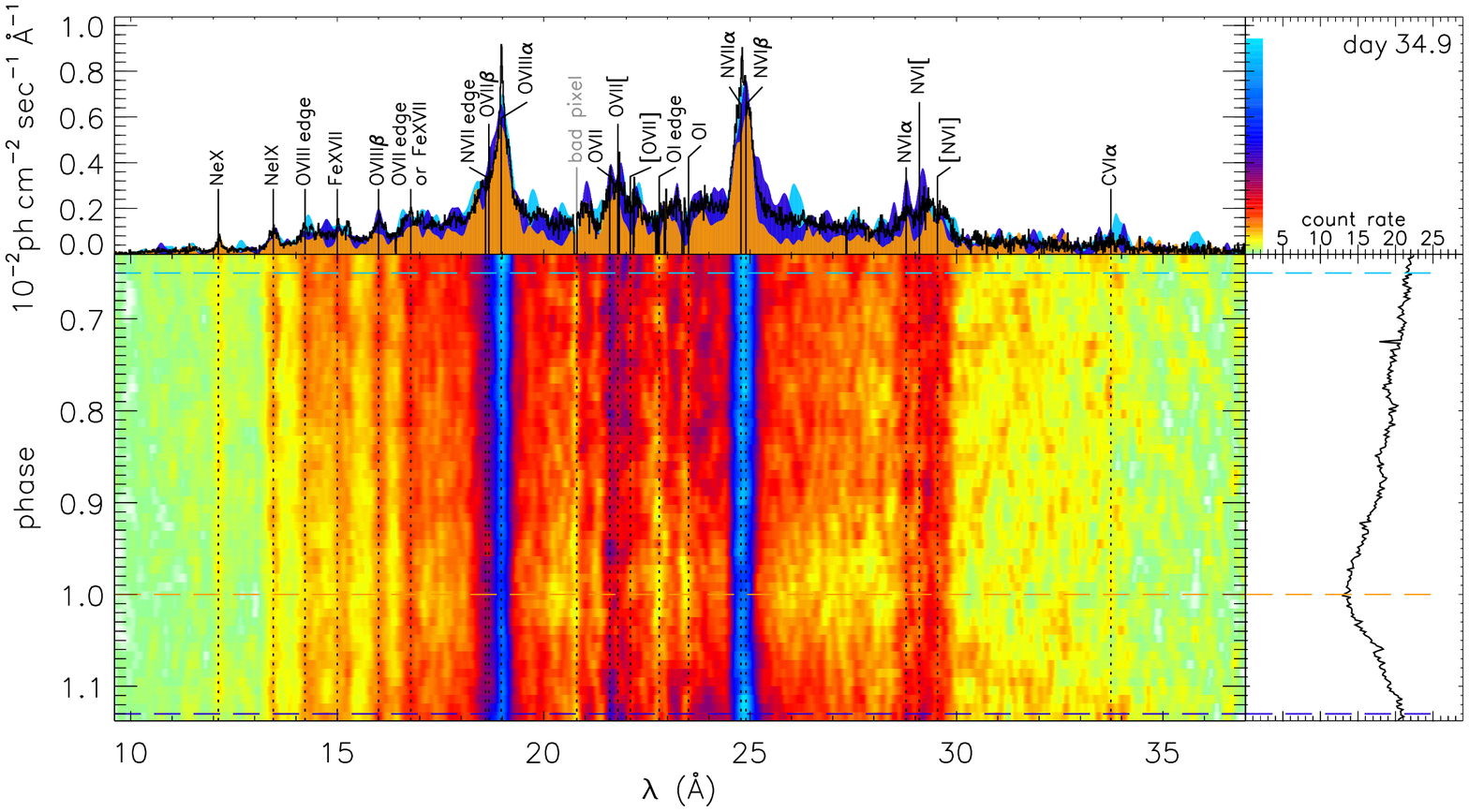}}
\caption{\label{smap}Illustration of time evolution of RGS spectra taken
on days 22.9 (top) and 34.9 (bottom). The central part in each plot is
a brightness map composed of 63 and 54 separate spectra (1000\,s exposure
time), respectively, with colors representing increasing flux levels from
light green,
yellow, red, purple, dark blue, and light blue. The corresponding
flux for each color can be determined using the vertical legend
bar in the top right. In the top panel, two isocontour levels are
included in white. Increasing wavelengths are plotted from left
to right, and increasing phase from top to bottom; phases
have been calculated from photon arrival times based on
Eq.~\ref{ephem}. The dashed horizontal lines running across the
brightness map indicate selected times for which the corresponding
spectra are shown with colored shades in the top panel.
In addition, the average spectra
from Fig.~\ref{bbfits} are included.
In the right, the pn light curves from the top panels of
Figs.~\ref{lc1} and \ref{lc2} are shown, rotated by
90$^{\rm o}$ clockwise. The dips in the top panel and the eclipse
in the bottom panel are seen in both continuum and emission lines
in the brightness map.
Flare-like events at phases 1.21 and 0.9 on day 22.9 yield an
increase in the N\,{\sc vi} lines at $\sim 29$\,\AA. The two
black lines illustrate a time lag between
flares and the changes in N\,{\sc vi} that correspond to
a light travel distance of 2\,AU.
}
\end{figure*}

{\bf B) Short term evolution}

 To study the short term evolution, we have extracted
multiple RGS and EPIC spectra covering adjacent time intervals.
In Figs.~\ref{smap} and \ref{smap_epic}, the spectra and
pn light curves, rotated by 90 degrees, are shown in the
top and right panels, respectively. The central parts are brightness maps
with increasing orbital phase from top to bottom and increasing
wavelength/energy from left to right for the RGS and EPIC data,
respectively. The key to the color scheme is given in the top right.
In addition, the hardness light curves shown in the second panels of
Figs.~\ref{lc1} and \ref{lc2} are used for the discussion.\\

To illustrate the effects of
contamination of the high-energy part of the EPIC spectra by solar
flare activity, field-integrated flare particle background light
curves are shown with light blue color in Fig.~\ref{smap_epic}.
To determine
the particle background, a full-field light curve was extracted above
10\,keV. For better comparison, the particle background light curve
is amplified by a factor 20.
 Above 2\,keV, no significant emission can be identified in the
EPIC spectra of either observation.
Around phase 1.3 in the first observation and phase 0.85 in the
second, some harder emission was present that could be contamination
because it coincides with elevated levels of solar flare particle background
emission.

 We have also extracted separate RGS spectra for times during dips
and peaks for day 22.9 and in- and outside of eclipse for day 34.9.
A combined dip spectrum has been extracted from the time intervals (in
spacecraft units $10^8$s after 1998-01-01) $3.8298385-3.8298715$,
$3.8298925-3.8299995$, and $3.8300645-3.8301275$.
A peak spectrum is integrated
over the five time intervals
$3.8298166-3.8298362$,
$3.8298726-3.8298878$,
$3.8300018-3.8300620$,
$3.8301286-3.8301858$, and
$3.8302500-3.8304502$.
 The inside eclipse spectrum was extracted from the
time interval $3.8404789-3.8407309$ and the outside eclipse spectrum
from $3.8401449-3.8403669$.
During extraction, the spectra were binned to 0.1-\AA\ bins,
larger than the default 0.01\,\AA\ for better signal to noise. The
differences between the resulting spectra are shown in
Fig.~\ref{diffspec}.
The shaded areas around the
difference spectra represent the uncertainty ranges derived
via error propagation from the measurement errors in the
individual spectra. If the times of low flux are interpreted
as occultations of emission originating from within the binary
orbit, the difference spectra are representative of the plasma
residing in the inner orbit while the low-flux spectra represent
only outer regions.\\ 

\begin{figure}[!ht]
\resizebox{\hsize}{!}{\includegraphics{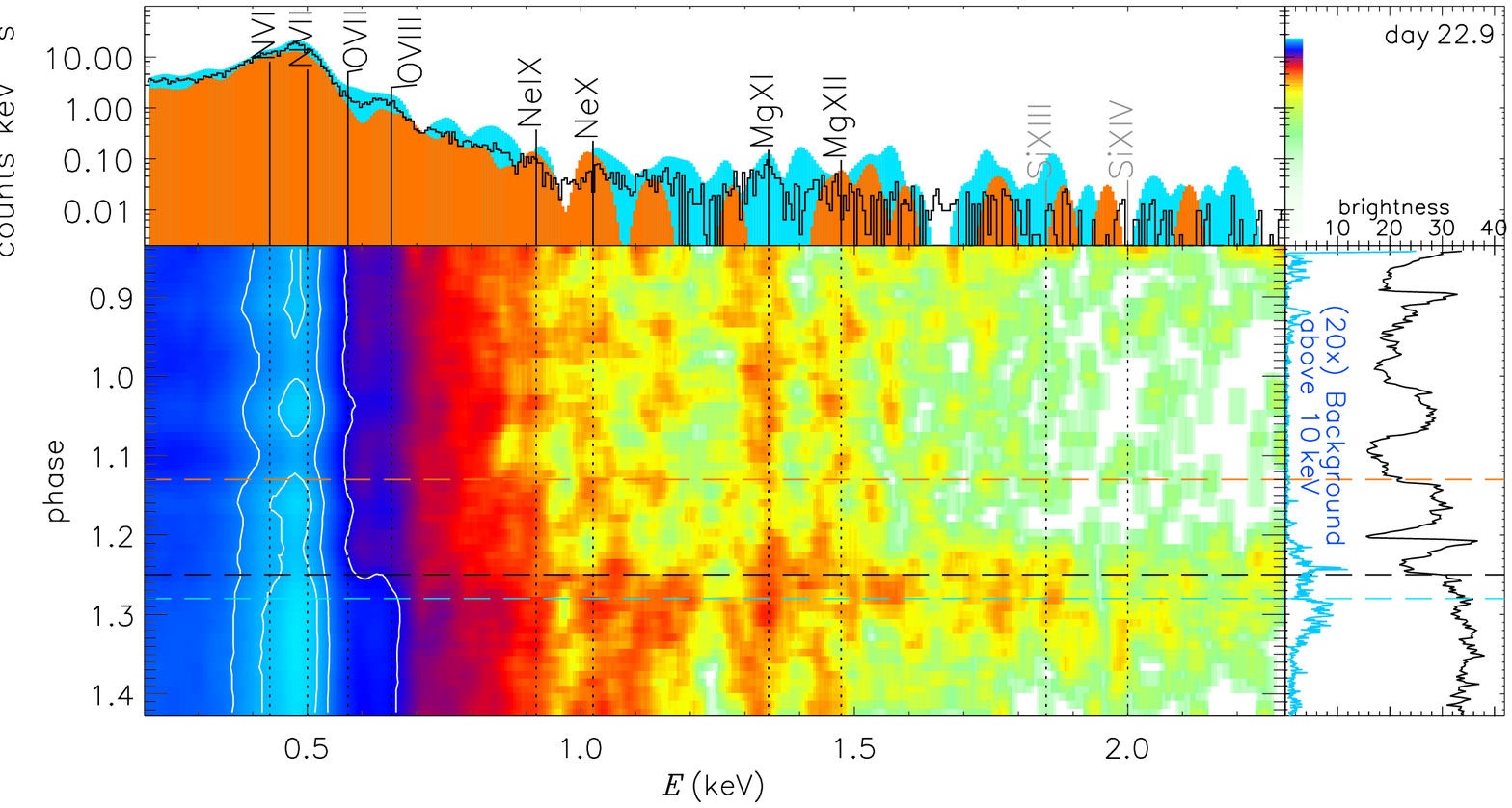}}

\resizebox{\hsize}{!}{\includegraphics{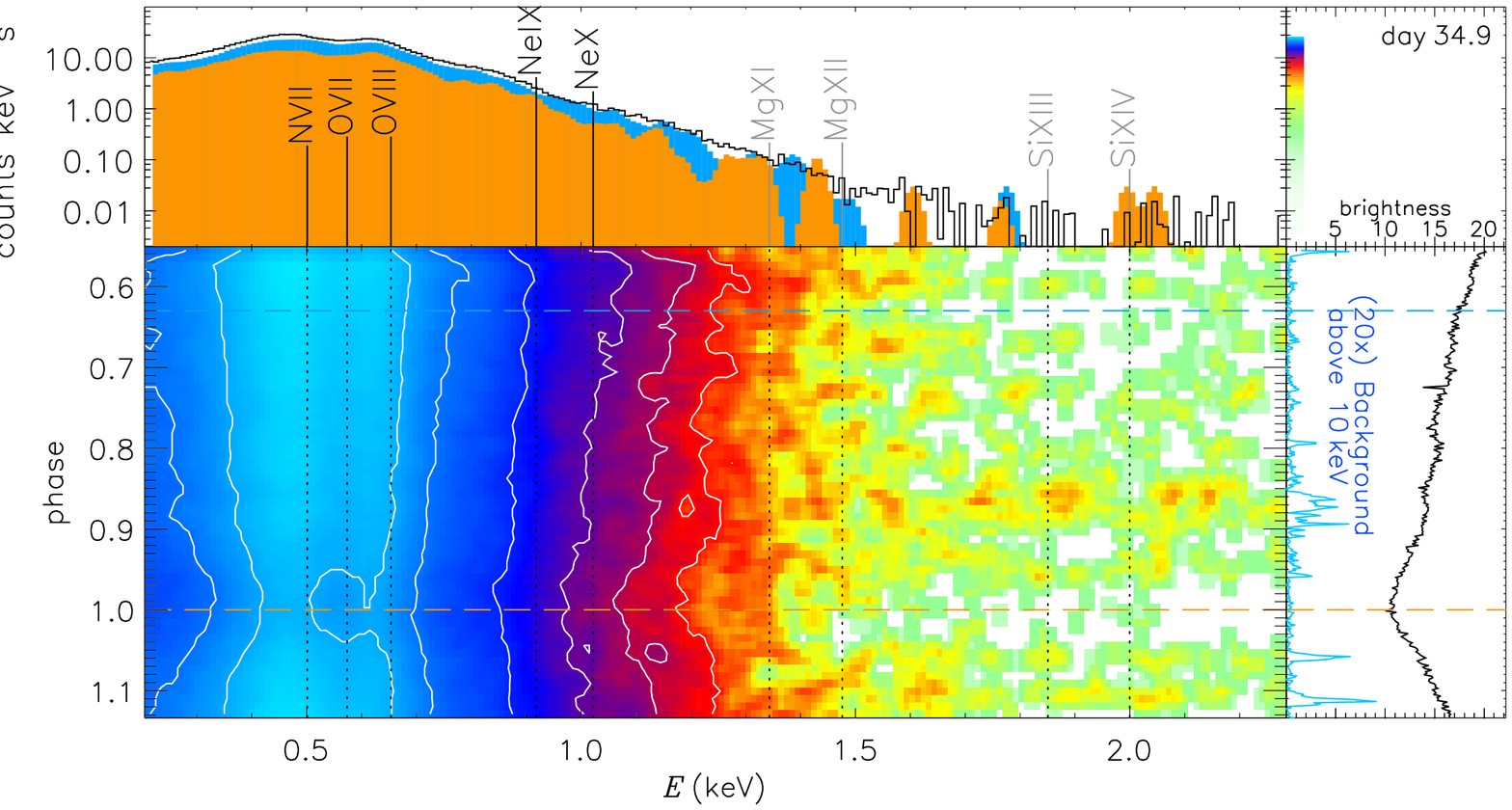}}
\caption{\label{smap_epic}Illustration of the spectral evolution in the
0.2-2.7\,keV band in the same format as explained in the legend to
Fig.~\ref{smap}, based on 63 and 62 EPIC spectra, extracted from adjacent time
intervals in the observations, starting day 22.9 (top, MOS2 detector)
and 34.9 (bottom, pn detector), respectively.
The white lines in the brightness maps are isocontours.
The energies of strong H-like and He-like transitions are marked in the
top parts (see also Fig.~\ref{epicspectra}). In the right, the same pn
light curve from the top panel of Fig.~\ref{lc1} is shown, rotated by
90$^{\rm o}$ clockwise. In addition, the field-integrated light curves
extracted from above 10\,keV are shown to illustrate times at which solar
flare activity may compromise the high-energy part of the spectra.
}
\end{figure}

{\bf B1) Flare-like events}

Two significant brightness increases occurred around phases
0.9 and 1.2 in the first observation which in the following
are referred to as flares. In the top panel of
Fig.~\ref{smap}, it can be seen in the contour levels that
both events were accompanied by increased emission in the N\,{\sc vi}
lines which originate from
lower ionization plasma than the other emission lines.
During the flare at phase 1.2, some extra emission
in the N\,{\sc vi} intercombination line can be seen. At the same
time, the N\,{\sc vii} Ly$\alpha$ line at 24.8\,\AA\
continues to be fainter during the dip while the higher
Ly series lines of N\,{\sc vii} seem to peak around the
same time as the  N\,{\sc vi} lines.

 The contemporaneous appearance of the lower-temperature
line with both flare-like events is unexpected because flares
are usually high-temperature events involving some heating
mechanism. For example, in solar and stellar coronae, flare
events are usually accompanied by the appearance of high-ionization
emission lines which are produced directly in the flare
\citep[e.g.,][]{guedel04}.
 
 When assuming photospheric events as origin of the
flares and that the N\,{\sc vi} line emission has
responded to these changes, then a time delay of $\sim 0.01$
in phase, corresponding to $\sim 17$ minutes after the peak
of each flare, can be used to derive the maximum distance
between the plasma that emits the excess in N\,{\sc vi} and
the photosphere. The two black horizontal dashed lines in
Fig.~\ref{smap} illustrate the time delay, and it corresponds
to a light travel distance of $\sim 2$\,AU. This is outside of
the binary orbit of 6.5\,R$_\odot$ where cooler plasma resides.
Hotter plasma further inside may have been blocked during the times
of the flares, as both occurred during dips.\\

{\bf B2) Spectral changes on day 22.9 (dipping)}

\begin{figure*}[!ht]
\resizebox{\hsize}{!}{\includegraphics{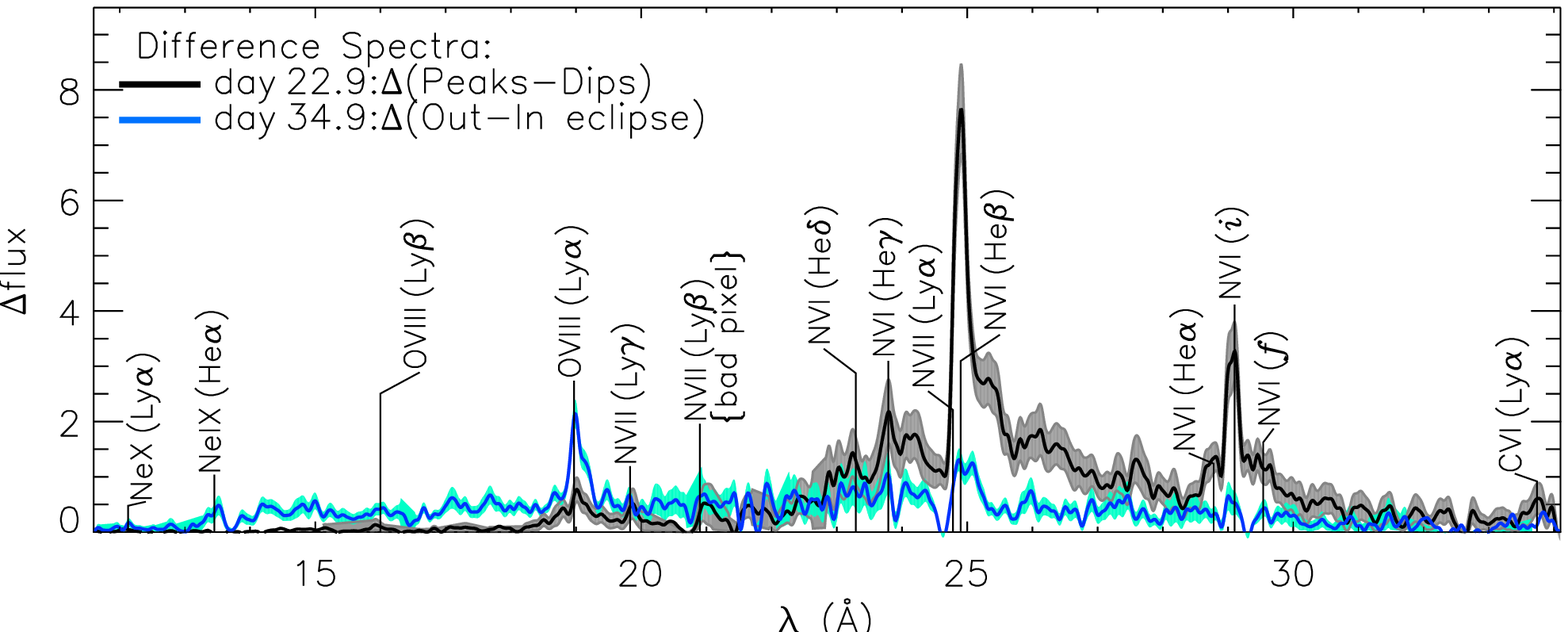}}
\caption{\label{diffspec}Difference spectra between
dip- and peak spectra for day 22.9 (gray) and in- and out of
eclipse spectra for day 34.9 (light blue), representing
the spectra of the missing emission during dip and eclipse,
respectively, thus from the inner-binary region.
The indices Ly$\alpha$, $\beta$, and $\gamma$ indicate
H-like Lyman series lines with 1s-2p to 1s-4p transitions,
and He$\alpha$ to $\delta$ are the respective He-like series lines
with transitions 1s-2p to 1-5p. The notation
He$\alpha$, $i$, and $f$ behind the N\,{\sc vi} labels indicate
He-like 1s-2p transitions with
$^1$S-$^1$P (resonance line),
$^1$S-$^3$P (intercombination line), and
$^1$S-$^3$S (forbidden line), respectively.
The N\,{\sc vi} intercombination line ($i$ at 29.1\,\AA)
and the He-like series lines $\beta$ to $\delta$ originate from
the inner orbit on day 22.9 and are thus photoexcited by the
nearby photospheric continuum emission. The N\,{\sc vi} He$\alpha$
line is reduced in the inner-orbit spectrum because of resonant line
scattering out of the line of sight. The O\,{\sc viii} resonance
line at 18.97\,\AA\ on day 34.9 reaches us more directly from
the inner regions while the N lines are reduced by the same
amount as the surrounding continuum.}
\end{figure*}

Closer inspection of the spectral changes with the dips in the
first observation gives insights into the geometry of the
emitting gas at this time. The brightness map in the top panel of
Fig.~\ref{smap} illustrates that during the dips, the continuum
and the emission lines are both fainter, but neither component
is completely occulted. Both components must therefore
originate from both, the inner orbit and from further outside.
Thompson scattering of the continuum component has already been
discussed in Sect.~\ref{rgsspec} to resolve the conflict between
high temperature from the Wien tail and the required small
size of the central source plus the absence of absorption lines.

 A comparison of the emission lines during and outside dips
on day 22.9 can be studied in Fig.~\ref{diffspec}.
The difference spectrum represents the
plasma within the inner orbit, revealing the spectral signatures
of resonance line scattering. While the strong 1s-2p resonance
lines of, e.g.,  N\,{\sc vii} (24.78\,\AA) and N\,{\sc vi}
(28.74\,\AA) originate only from the outer regions, lines
belonging to transitions involving higher principal
quantum numbers are stronger in the difference spectrum, e.g.,
the  N\,{\sc vi} He-like series lines He$\beta$, $\gamma$, and
$\delta$. The respective oscillator strengths are $f=0.17$,
$0.05$, and $0.02$, which are low compared to the oscillator
strength of the He$\alpha$ line of $f=0.71$. Line photons with
high values of $f$ can be absorbed and re-emitted in a different
direction. In a non-spherical plasma, strong resonance lines
can be amplified or reduced, compared to lines of lower
oscillator strengths. The fact that we are seeing the
high-$f$ lines from the inner regions reduced is consistent
with the geometry of an accretion disk. 

 Forbidden transitions can not be absorbed, and the
contributions from the inner region to the ratio of
the N\,{\sc vi}
intercombination and forbidden lines at 29.1 and 29.54\,\AA\
is unaltered by resonance scattering. Owing to the 
proximity to the central source, the low f/i ratio reported
in Sect.~\ref{rgsspec} has to be interpreted as
photoexcitation rather than high density.

 In the top panel of Fig.\ref{smap_epic}, the evolution of the
Wien tail can be studied that is a qualitative indicator for
the photospheric temperature. During dips, the continuum
emission has gone down at all wavelength, and no shift of
the Wien tail can be seen that would indicate that the
continuum component from the outside regions has a
different temperature. This supports the interpretation
of Thompson scattering. The sudden change in the variability
pattern at
phase 1.25 is accompanied by a sudden increase in brightness
above 0.6\,keV, extending to $\sim 1.5$\,keV, including the
Mg\,{\sc xi} 1s-2p transition at 1.3\,keV. While solar flare
activity could be held responsible for the brighter emission
between phases 1.2-1.35, the spectra taken after phase 1.35 are
not contaminated and are still harder. This could be interpreted
in the general context of increasing temperature on longer
time scales as discussed above. Also, a slight
increase in the high-ionization O\,{\sc viii} line is suggestive
in the top panel of Fig.~\ref{smap}, but it may also be
part of the general increase in brightness.\\

{\bf B3) Spectral changes on day 34.9 (eclipse)}

 The spectral evolution during the second observation is
displayed in the bottom panels of Figs.~\ref{smap} and
\ref{smap_epic}. The continuum and the emission line components
experience similar changes during the eclipse as during the dips
on day 22.9. The total emission decreases by $\sim 50$\% during
eclipse, indicative of significant scattering processes.
The emission lines decrease by roughly the same amount as
the continuum.

 This is illustrated with the difference spectra
shown with blue color in Fig.~\ref{diffspec}.
The N\,{\sc vi} and N\,{\sc vii} lines
are hardly present, and therefore, {\em all} the plasma emitting
the observed N lines resides outside of the binary orbit
while some of the flux in the hotter O\,{\sc viii} line also
originates from within the orbit. In order to explain the changes
from day 22.9 to 34.9, the line emission that was first seen to
come from the inner regions (N\,{\sc vi} He$\beta$ and
intercombination line) has to be moved to further outside to
escape eclipses by the companion on day 34.9. While for the
N\,{\sc vi} He-like intercombination line one could argue
that the entire N\,{\sc vi} triplet is too faint for studies
of changes, the line complex around 25\,\AA\ is strong
enough that residual N\,{\sc vi} He$\beta$ would be
detectable. Also, the higher Ly series lines of N\,{\sc vii}
are not seen in the difference spectrum. We interpret the
absence also of high Ly series lines in the difference
spectrum as evidence for an increase in the optical depth
of the scattering plasma, which could be part of the
reformation process of the accretion disk. A bit puzzling
appears the presence of O\,{\sc viii} Ly$\alpha$ in the
difference spectrum while Ly$\beta$ is not present.


Except for the O\,{\sc viii} Ly$\alpha$ line, the reduction
in brightness during eclipse can therefore exclusively be attributed
to the continuum. The shape of the continuum in the difference
spectrum in Fig.~\ref{diffspec} is the shape of the continuum
component in the inner regions. It has roughly the same shape
as the out-of-eclipse spectrum, supporting the interpretation
of achromatic Thompson scattering.

 Close inspection of the RGS and EPIC brightness maps
in the bottom panels of Figs.~\ref{smap} and \ref{smap_epic} 
reveals that the eclipse progresses in a slightly non-uniform
way in wavelength/energy. In the Rayleigh-Jeans tail, between
26-28\,\AA, the continuum seems to go through a wider eclipse
towards longer wavelength (Fig.~\ref{smap}). In the Wien tail,
the eclipse seems to be narrower towards higher energies
which is indicated by the contours in Fig.~\ref{smap_epic}.
This could indicate a temperature gradient, however, the
hardness light curve in the middle panel of Fig.~\ref{lc2}
does not indicate a significant temperature change with the eclipse.
Only the softening after the eclipse appears noteworthy and may
have to be attributed to the cooling while nuclear burning is turning
off. On the other hand, the hardness light curve contains the
emission lines and is a less sensitive indicator for the
photospheric temperature.

\subsection{UV spectra}
\label{uvspec}

\begin{figure*}[!ht]
\resizebox{\hsize}{!}{\includegraphics{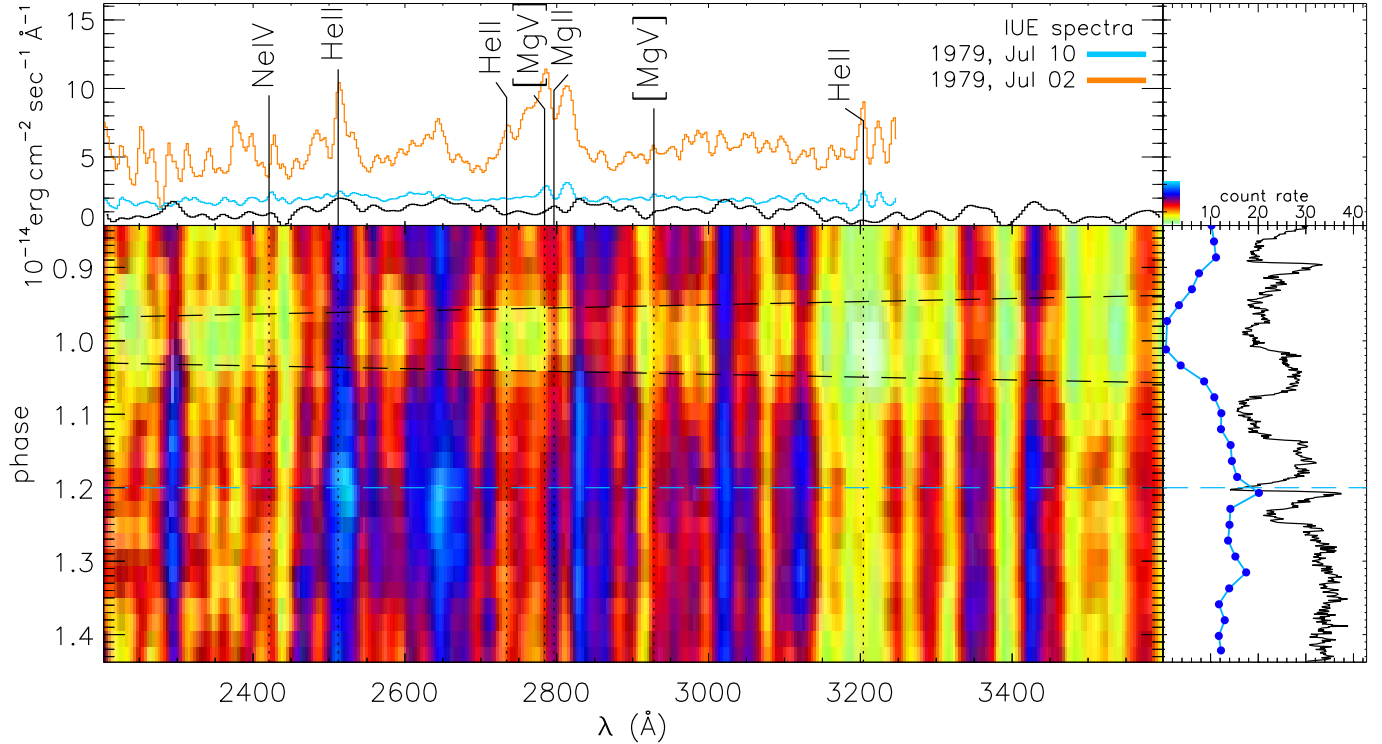}}
\caption{\label{smap_om}Illustration of the spectral evolution in
the same format as Fig.~\ref{smap} for 27 OM U grism observations,
starting on day 22.9. In the top panel, two IUE spectra taken 8 and
16 days
after the 1979 outburst of U\,Sco are included in orange and blue
as indicated in the right legend.
In the right panel, the X-ray light curve and the OM light curve
are shown in black and blue, respectively. The two diverging black
dashed lines in the central panel are drawn to guide the eye,
indicating a possible trend of broader eclipse towards longer
wavelengths. The light blue dashed line indicates the time of the
flare at phase 1.2.
}
\end{figure*}

In the first of the two \xmm\ observations, starting day 22.9,
the UV grism of the OM was employed, taking 27 consecutive spectra
(see Table~\ref{obslog}). In Fig.~\ref{smap_om}, these spectra are
shown in the same format as in Figs.~\ref{smap} and \ref{smap_epic}.
In the range 2200-3600\,\AA, the spectrum shows
many features which are difficult to identify. Wavelengths of a few
known lines are marked. A weak
feature coincides with the position of the 2800\,\AA\ Mg\,{\sc ii} resonance
doublet but, given the presence of many similar features in the
spectrum as well as the uncertainty of the absolute wavelength scale
(which depends on the accurate centroiding of the zero order), its
identification is uncertain.

 The wavelength range of the OM grisms overlaps with the long-wavelength
range of the International Ultraviolet Explorer (IUE). Five spectra
were taken with the IUE during the outburst in 1979: on June 28,
June 30, July 2, July 4, and July 10, corresponding to 5, 7, 8, 10, and 16
days after outburst \citep{williams81}. In the top panel of
Fig.~\ref{smap_om}, the IUE spectra from 8 and 16 days after the 1979
outburst are overplotted in orange and blue (see right legend). The
strongest lines in these spectra are, apart from the
Mg\,{\sc ii} doublet, He\,{\sc ii} at 2512, 2734 and 3204\,\AA,
Ne\,{\sc iv} at 2421\,\AA\ and [Mg\,{\sc v}] at 2784 and 2928\,\AA.

The eclipse minimum can be seen to coincide with reductions in flux at
all wavelengths in the brightness image, but some spectral ranges,
e.g., $\sim 2350$\,\AA, yield narrower eclipses than others, e.g.,
$\sim 3200$\,\AA. The two dashed black lines in the image draw attention to the
possibility of broader eclipses towards longer wavelengths as found in
\S\ref{lcmodel} for the two wavelength bands 2200-2800\,\AA\ and
2800-3600\,\AA. If
these bands are interpreted as spectral colors, the conclusion would
be that hotter plasma resides further inside. However, the concept of
spectral color as temperature indicator is only valid for continuum
sources while emission lines can arise at any wavelength,
independently of their formation temperature. While emission lines of
higher ionization stages are produced closer to the central ionizing
source, they do not necessarily arise at shorter wavelengths. It can
be seen that some features, e.g., at 2650\,\AA, have indeed narrower
eclipse profiles than others, e.g., at 3125\,\AA. Also, not all features
have their eclipse minimum at the same phase, suggesting that, if they
are real lines, they arise from different emission regions in an
inhomogeneous plasma.

\section{Discussion}
\label{disc}

 The different behavior in the simultaneous \xmm\ X-ray and UV light
curves starting on day 22.9 was unexpected with a clear eclipse only in UV
and optical but multiple dips in the X-rays. Some similarities with
the class of X-ray dippers \citep{veryfirst_dipper,first_dipper} may
be helpful in discussing possible scenarios. For example, \cite{trigo09}
presented in their figure 5 the \xmm\ X-ray and UV light curves of
the high-inclination X-ray binary XB\,1254-690 and found clear eclipses
in the V band while in the X-ray band, deep dips were seen at phase 0.8. 
While multiple dips are generally not seen in the X-ray dippers,
EXO 0748-676 contained dips at phases 0.6, 0.7, 0.8, 0.9 and 0.1
(see figure~9 in \citealt{parmar86}). While a number of
similarities are thus seen, U\,Sco is still a completely different
system, and any parallels to X-ray dippers have to be interpreted
with care.\\

In X-ray dippers, significant changes in spectral X-ray hardness have
oftentimes been observed, while in U\,Sco, the hardness remained the
same during dips. An achromatic steep decrease in flux has also been
seen by \cite{sala08}in the nova V5116\,Sgr.
In contrast to U\,Sco and other novae, the central object in X-ray
dippers is an accreting neutron star, emitting a much harder continuum.
Harder X-rays can partially penetrate material of large column densities,
and only the soft tail of the X-ray spectrum is blocked. The continuum
spectrum of U\,Sco is much softer, and any material of significant
column density above $\sim 10^{21}$\,cm$^{-2}$ will block all emission
from behind, regardless of energy. Total disappearance of all X-ray
light during dips is only avoided because of Thompson scattering of the
central emission into a much larger region rather than partial
transparency of the absorbers. Our parameter studies of $N_{\rm H}$ in
the light curve model support this interpretation (see Fig.~\ref{nh_chi}).
An example of achromatic dipping at low energies as a consequence of
high-$N_{\rm H}$ absorbers in relation to the shape of the source
spectrum is Cyg X-2 \citep{cygx2}. While the continuum is much harder
than in U\,Sco, the opacity of the absorbers is
$\sim 10^{23}$\,cm$^{-2}$, leading to full blockage of all light in
the 0.5-2\,keV band. The hardness ratio from 0.5-1.0/1.0-2.0 band is
therefore constant during the dips, while only emission from the
4.5-8.0\,keV band can partially penetrate. In U\,Sco, not enough
emission at energies above 1\,keV is present to constrain
$N_{\rm H}$ of the absorbers in the way this is possible for Cyg X-2.\\

 A question of interest is the reformation of the accretion disk
that was destroyed during the nova outburst \citep{drake10}.
The transition from dipping on day 22.9 to clean eclipse on
day 34.9 may be part of the reformation process, and the concept
of intersecting opaque clumps as part of the reformation process
deserves closer consideration. The theoretical possibility of clumps in
the ejecta has been discussed by \cite{diaz10}, however, being immersed
in a high-radiation environment, the ejecta must be highly ionized, and
as such, they do not absorb soft X-rays via the photoelectric effect.
Thompson scattering in clumpy material can also be ruled out as cause
for the dips as that would also lead to the same dips in the
UV/optical light curves. Clumpy ejecta
can therefore not cause such deep dips in the X-ray light curve.
A possible source of opaque clumps consisting of cooler material
is the companion star, if accretion has set in again. Cooler plasma
is opaque to soft X-ray while at the same time being transparent
to light with wavelengths longer than 912\,\AA\ (13.6\,eV), the
ionization energy of atomic hydrogen, which would explain why no
dips are seen in the simultaneous UV and optical light curves.
Absorbing atomic material from the companion is thus strong
support for the interpretation that accretion has set in again.
This does not require the full reestablishment of the accretion
disk, a clumpy accretion stream can also be imagined. The absence of
dips after phase 1.25 suggests that no clumpy material is on the far
side as seen from the companion which is an argument in favor of an
accretion stream, residing between the companion and the central
white dwarf.\\

\begin{figure}[!ht]
\resizebox{\hsize}{!}{\rotatebox{270}{\includegraphics{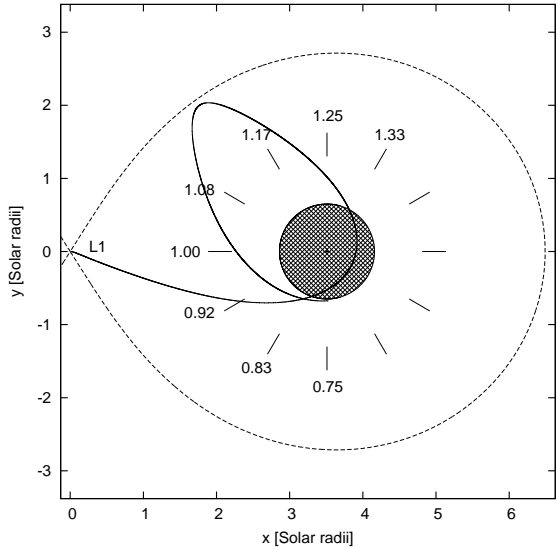}}}
\caption{\label{trajectory}Trajectory of accretion stream at an early time of the reformation of
the accretion disk. The calculation follows \cite{Flannery75_mnras} using
the parameters of U\,Sco. The trajectory interacts with itself, forming an
asymmetric ring at the beginning of disk reformation. The radius of the
central source is drawn assuming the radius derived from the light curve
model (see top panel of Fig.~\ref{emap22}), which is also consistent with
the blackbody radius. The trajectory dives into the optically
thick ejecta on the far side from the inner Lagrangian point (L1). This
could explain the residual variability seen in the bottom panel of
Fig.~\ref{emap22}. Being inside the
photosphere, this part of the accretion stream does not occult
the photospheric X-ray emission. No dips are therefore expected between
phases 1.33 and 0.75, while during other times, dips can occur, depending
on the structure of the accretion stream. At a later time, the circular
ring that is the protodisk interacts with the incoming stream, leading
to circularization that could lead to the disappearance of dips in
X-rays as seen on day 34.9 (see Fig.\ref{lc2}).
}
\end{figure}

In Fig.~\ref{trajectory}, we show a theoretically calculated initial
trajectory of a reforming disk based on equations 1 and 2 in
\cite{Flannery75_mnras}, assuming the parameters of U\,Sco.
During the early disk formation phase, an asymmetrical
"ring" is formed. The phases that correspond to different
viewing angles are marked. The hashed area in the center
encircles the optically thick parts of the ejecta. The surface
can be considered the source of X-ray emission, and no views
inside are possible. The highly elliptical
trajectory runs through the optically thick part of the
ejecta, and any inhomogeneities inside the photosphere
can not cause dips. An eccentric trajectory of infalling
material can thus be the explanation for not seeing any
dips after phase 1.25.\\

 A remarkably similar event has been observed by \cite{zand2011}
in the accreting neutron star in 2S 0918-549. After an X-ray burst,
strong ($87\,\pm\,1$\% peak-to-peak amplitude) achromatic fluctuations
were seen as a transient phenomenon. Since the outburst evolved
much more rapidly, the X-ray light curve taken with the Proportional
Counter Array (PCA) on the Rossi X-ray Timing Explorer (RXTE) may
be interpreted as a fast-motion picture of the evolution in U\,Sco
(see figure 1 in \citealt{zand2011}).
It was speculated by \cite{zand2011} that the fluctuations were due to
Thompson scattering by inhomogeneities in a resettling
accretion disk that was disrupted by the effects of super-Eddington
fluxes, much like in U\,Sco where the blast wave has destroyed the
accretion disk \citep{drake10}. Furthermore, they argue that an
expanding shell may be the necessary prerequisite for the
fluctuations. Irregularities in the accretion
stream could be caused by the continued expansion of the nova ejecta
that has stopped between days 22.9 and 34.9. The stream could be
disrupted by the ejecta, or the companion star may be distorted in
the outer layers. A difference to \cite{zand2011} is the
interpretation of the properties of the intervening material,
yielding highly ionized plasma. Dips are caused by Thompson
scattering out of the line of sight rather than photoelectric
absorption. This is consistent with
their observation of increases above the decay trend of the burst
which can be explained by sideways or backward scattering of clouds
located behind the compact object. Such increases are not seen in
U\,Sco. Furthermore, achromatic Thompson scattering would also
leave its footprints in the UV and optical light curves, and
photoelectric absorption by high-$N_{\rm H}$ plasma in a low state
of ionization appears more likely for U\,Sco.

\section{Summary and Conclusions}

 From the simultaneous \xmm\ X-ray, UV, and optical observations
of the 10th recorded outburst of the recurrent, eclipsing nova
U\,Sco, we have deduced new insights into time variability, 
spectral properties, and spatial structure:
\begin{itemize}
\item {\bf Time variability}: As opposed to expectations that
eclipse mapping will reveal the photospheric radius of the
white dwarf in different stages of the evolution, we have seen
achromatic dipping in X-rays but clean eclipses in UV and
optical. Dips were only seen before quadrature, phase 1.25.
The
later observation contains clean eclipses in all bands. During
dips and eclipses, $\sim 50$\% of X-ray emission remained, indicating
that the observed X-ray emission is extended beyond the binary orbit.

\item {\bf Spectral Properties}:
 The high-resolution X-ray spectra consist of blackbody-like
continuum emission plus strong emission lines. No
absorption lines can be identified. From day 22.9 to day 34.9,
an increase in temperature can be deduced from the
shape of the continuum and from increases in emission lines
arising in ions of higher ionization stages.

 The eclipsing supersoft source Cal\,87 yields a qualitatively
similar
high-resolution X-ray spectrum. Photospheric X-ray emission from the
central source is blocked by the accretion disk and can only
be seen after Thompson scattering in the accretion disk corona.
Strong emission lines come from resonant line scattering within
the surrounding optically thin medium. Only emission lines from
transitions with low oscillator strengths are seen to originate
directly from the inner regions.

The UV grism spectra on day 22.9 contain emission lines that
undergo eclipses, which indicates that they originate
from within the inner binary orbit. 
The interpretation as an emission
line spectrum is supported by a comparison to the 1979 outburst,
during which the UV spectra were already optically thin
7 days after outburst.

\item {\bf Evolution of the Spatial Structure}: We interpret the
dipping as the
result of occultations of the central source by high-density
absorbing gas that is aligned along the trajectory of a reforming
accretion stream. The gas consists of plasma of low ionization
stage, originating from the companion star. The continued outflow 
disrupts the accretion stream into fragments that intersect the line
of sight. After nuclear burning has ceased, the outflow stops, no more
fragmentation takes place, allowing the accretion stream to circularize
and condense into a flat accretion disk, yielding no more dips.
\end{itemize}

 One important argument in favor of reestablishment of accretion
is the low degree of ionization of the intervening material which
indicates that it originates from the companion. The most important
supporting observations are the achromatic nature of dips and the
lack of dips in the other
wavelength bands. Ionized plasma would not cause any photoelectric
absorption while low-ionization plasma is only opaque to
soft X-rays but transparent to optical/UV emission,
longward of 912\,\AA. While the dips could be caused by Thompson
scattering in the electron-rich environment of the highly ionized
nova ejecta, the same dips would have to be seen in UV and optical.\\

The observed spectral signatures for combined Thompson scattering
and resonant line scattering are:
\begin{itemize}
 \item The continuum component does not go to zero during either dips
   or eclipse. The location of the Wien tail requires the
   source to be small enough for complete occultations to avoid
   super Eddington luminosity. Achromatic Thompson
   scattering can increase the geometric size of the
   emission region without changing the shape of the
   continuum. Photospheric absorption lines are smeared out
   in the fast-moving scattering plasma and are therefore
   not resolved.
 \item Our diagnostics of dip and non-dip spectra revealed that
   only emission lines from transitions with low oscillator
   strength originate directly from the inner regions. During dips,
   the flux in strong resonance lines is lower by the same
   amount as the surrounding continuum, indicating that
   they are scattered out of the line of sight, leaving us
   to only see them after resonant scattering.
\end{itemize}

The observed history of the reformation of the accretion disk
in U\,Sco is likely applicable to formation of accretion disks
in general, although the time scales may depend on
system parameters, which bears closer investigation via
theoretical simulations.

\acknowledgments

 Based on observations obtained with \xmm, an ESA science mission
with instruments and contributions directly funded by ESA Member States
and NASA.
We'd like to thank the \xmm\ Science Operations Centre for
efficient scheduling and generously granting Director's Discretionary
Time for the second observation. B.E.S. was supported under a grant
from the National Science Foundation (AST 0708079).
AD was supported in part by the Grand-in-Aid for the global COE programs
on "The Next Generation of Physics, spun from Diversity and Emergence"
from MEXT and also by the Slovak Grant Agency, grant VEGA-1/0520/10.
KLP acknowledges support from STFC.
JJD was  supported by NASA contract
NAS8-39073 to the {\em Chandra X-ray Center}.
SS acknowledges support from NASA+NSF grants to ASU.
MH acknowledges support from MICINN project AYA 2008-01839/ESP,
AGAUR project 2009 SGR 315 and FEDER funds and GS MICINN projects AYA
2008-04211-C02-01 and AYA 2007-66256.
GS acknowledges MICINN grants AYA2008-04211-C02-01 and AYA2010-15685, and ESF EUROCORES Program EuroGENESIS grant EUI2009-04167.

\bibliographystyle{apj}
\bibliography{cn,usco,astron,jn,rsoph}

\end{document}